\documentclass[12pt,letterpaper]{article}  
\usepackage[includeheadfoot,
            marginratio={1:1,2:3}, 
            width=412pt, 
            height=688pt,]{geometry}

\usepackage{amsmath}
\usepackage{amsfonts}
\usepackage{amssymb}
\usepackage{mathtools}
\usepackage{bbold}
\usepackage{graphicx,caption,subfigure}
\usepackage{tikz} 
\usetikzlibrary{decorations.pathreplacing,calc}
\usepackage{booktabs}
\usepackage{adjustbox}
\usepackage[utf8]{inputenc}
 \usepackage[splitrule]{footmisc} %% The splitrule option draws a full width rule above the continued part of the footnote as a visual cue to readers.

\usepackage{empheq}
\usepackage{paralist}
\usepackage{cite}
\usepackage[normalem]{ulem}

\usepackage[colorlinks=false,urlbordercolor=red]{hyperref}
\usepackage{xcolor}
\usepackage{tensor}

\newcommand{\nc}{\newcommand}
\nc{\lb}{\llbracket}
\nc{\rb}{\rrbracket}
\nc{\gl}{\llbracket}
\nc{\gr}{\rrbracket}
\nc{\del}{\partial}
\nc{\tri}{\hspace{-3.5pt}\vartriangle\hspace{-3.5pt}}
\nc{\blacktri}{\blacktriangle}

\nc{\eq}[1]{\begin{equation}
                     \begin{split} #1 \end{split}
                     \end{equation}}
\nc{\ul}{\underline}
\nc{\ov}{\overline}

\nc{\fa}{\hat}
\nc{\fb}{\MakeUppercase}
\nc{\fc}{\tilde }
\nc{\Lie}{{\cal L}} 
\nc{\lambdabar}{{\mkern0.75mu\mathchar '26\mkern -9.75mu\lambda}}

\allowdisplaybreaks[2]
\numberwithin{equation}{section}

\DeclareUnicodeCharacter{2212}{-}
\begin{document}

\vspace*{-1.5cm}
\begin{flushright}
  {\small
  MPP-2023-33\\
  }
\end{flushright}

\vspace{1.5cm}
\begin{center}
  {\Large
    Dynamical Cobordism Conjecture:\\  Solutions for End-of-the-World  Branes \\[0.3cm]
} 
\vspace{0.4cm}

\end{center}

\vspace{0.35cm}
\begin{center}
Ralph Blumenhagen, Christian Knei\ss l and Chuying Wang
\end{center}

\vspace{0.1cm}
\begin{center} 
\emph{
Max-Planck-Institut f\"ur Physik (Werner-Heisenberg-Institut), \\[.1cm] 
   F\"ohringer Ring 6,  80805 M\"unchen, Germany } 
   \\[0.1cm] 
 \vspace{0.3cm} 
\end{center} 

\vspace{0.5cm}

%%%%%%%%%%%%%%%%%%%%%%%%%%%%%%%%%%%%%%%%%%%%%%%
%%%%%%%%%%%%%%%%%%%%%%%%%%%%%%%%%%%%%%%%%%%%%%%
%%%%%%%%%%%%%%%%%%%%%%%%%%%%%%%%%%%%%%%%%%%%%%%
%%%%%%%%%%%%%%%%%%%%%%%%%%%%%%%%%%%%%%%%%%%%%%%
%%%%%%%%%%%%%%%%%%%%%%%%%%%%%%%%%%%%%%%%%%%%%%%
%%%%%%%%%%%%%%%%%%%%%%%%%%%%%%%%%%%%%%%%%%%%%%%
%%%%%%%%%%%%%%%%%%%%%%%%%%%%%%%%%%%%%%%%%%%%%%%
%%%%%%%%%%%%%%%%%%%%%%%%%%%%%%%%%%%%%%%%%%%%%%%

\begin{abstract}
  We analyze finite size solutions for a generalized $D$-dimensional 
  Dudas-Mourad (DM) model featuring dynamical cobordism
  with neutral and charged end-of-the-world (ETW) defect branes.
   Confirming  a dynamical version of the  Cobordism
  Conjecture, we explicitly construct    non-isotropic solutions
  for the latter codimension one branes and show the appearance of a
  lower bound $\delta\ge 2\sqrt{(D-1)/(D-2)}$
  for the critical exponent in the scaling behavior
  of the distance and the curvature close to the wall. 
    This allows us to make a connection to the 
    (sharpened) Swampland Distance Conjecture and
    the (Anti-) de Sitter Distance Conjecture.
    Moreover, BPS orientifold planes appear as  special cases in our analysis
    and  the whole picture is consistent
    with dimensional reduction from ten to $D$ dimensions.
   An analogous analysis is performed for a generalized
   Blumenhagen-Font (BF) model featuring neutral codimension two
    ETW-branes where the same lower bound for the scaling parameter $\delta$  arises. 
    \end{abstract}

\clearpage

%\tableofcontents

%%%%%%%%%%%%%%%%%%%%%%%%%%%%%%%%%%%%%%%%%%%%%%%
%%%%%%%%%%%%%%%%%%%%%%%%%%%%%%%%%%%%%%%%%%%%%%%
%%%%%%%%%%%%%%%%%%%%%%%%%%%%%%%%%%%%%%%%%%%%%%%
%%%%%%%%%%%%%%%%%%%%%%%%%%%%%%%%%%%%%%%%%%%%%%%
%%%%%%%%%%%%%%%%%%%%%%%%%%%%%%%%%%%%%%%%%%%%%%%
%%%%%%%%%%%%%%%%%%%%%%%%%%%%%%%%%%%%%%%%%%%%%%%
%%%%%%%%%%%%%%%%%%%%%%%%%%%%%%%%%%%%%%%%%%%%%%%
%%%%%%%%%%%%%%%%%%%%%%%%%%%%%%%%%%%%%%%%%%%%%%%

%\newpage

\section{Introduction}
\label{sec:intro}

The incompatibility of global symmetries and quantum gravity \cite{Banks:2010zn,Banks:1988yz} is often
regarded as one of the most solid and pivotal swampland conjectures \cite{Palti:2019pca,vanBeest:2021lhn,Grana:2021zvf}.
Unfortunately, due to its universal formulation it is very hard to extract meaningful phenomenological restrictions from this conjecture.
Nevertheless, extensions of this notion have been proposed in the last couple of years and have been proven to be vital for increasing our understanding of the mathematical backbone of quantum gravity. 
For example topological global symmetries have a very rich mathematical description in terms of cobordism. The requirement of trivializing the cobordism group associated to quantum gravity \cite{McNamara:2019rup} follows directly from the absence of global symmetry in quantum gravity. 
To finally obtain a trivial cobordism group the global symmetries can either be broken by defects carrying appropriate topological charge 
or gauged by the introduction of a gauge field.
There has been a lot of progress in exploring the far reaching consequences of the Cobordism Conjecture recently. 
We would like to refer the reader to \cite{GarciaEtxebarria:2020xsr,Montero:2020icj,Dierigl:2020lai,Hamada:2021bbz,Debray:2021vob,McNamara:2021cuo,Blumenhagen:2021nmi,Andriot:2022mri,Blumenhagen:2022bvh,Velazquez:2022eco,McNamara:2022lrw,Dierigl:2022reg,Debray:2023yrs} for more insight on the recent developments regarding the Cobordism Conjecture.

While the Cobordism Conjecture itself is a statement about the topology required by quantum gravity, in a series of papers so called dynamical cobordisms were explored \cite{Buratti:2021yia,Buratti:2021fiv,Angius:2022aeq,Blumenhagen:2022mqw,Angius:2022mgh}. 
Instead of an apparent topological inconsistency,
they were investigating the effective supergravity description of theories with dynamical tadpoles, i.e. a running scalar potential breaking maximal symmetry. 
These rolling solutions share the common feature that they exhibit a
singularity at finite space-time distance, where also  a scalar   goes
to  infinite distance in field space\footnote{We note that such a feature was  described in
\cite{Antonelli:2019nar} under the name of ''pinch-off"-singularity.}. In the same way, as
a non-vanishing cobordism element can be made null-cobordant
by adding co-dimension one defects,   this  singularity at finite
distance signals the required presence of an end-of-the world (ETW) brane,
where space-time just ends.

While some scaling properties of such ETW-branes
can be inferred from the behavior of the running solution
\cite{Buratti:2021yia,Buratti:2021fiv,Angius:2022aeq} close to the
singularity, it was argued in
\cite{Blumenhagen:2022mqw} that this ETW-brane itself is  a local solution of the corresponding
equations of motion, as well. This was explicitly exemplified
for the rolling solution of the Blumenhagen-Font \cite{Blumenhagen:2000dc}  model. 
In this paper, we will elaborate more on this picture and convey 
a dynamical version of the Cobordism Conjecture:
For a  solution of an effective (super-)gravity theory
featuring a singularity at finite space-time distance, consistency
with quantum gravity requires  the 
existence of  an explicit solution for the  End-of-the-World (ETW) brane that
closes off the space-time.
In particular, this means that an appropriate ETW-brane solution of the EFT has to
match precisely the  scaling of the rolling solution  close to the
singularity. Since in most cases the configurations are
non-supersymmetric, we are not (yet) making any assumptions or claims 
about stability of the solutions.\footnote{For recent work on the issue of stability of closely related setups we would like to refer the reader to \cite{Basile:2021vxh, Raucci:2022bjw}.}

Analogous to  \cite{Basile:2022ypo}, we first systematically study a $D$-dimensional
Dudas-Mourad-like model  with $D\ge 3$ and a defining  exponential potential $V\sim
\exp(c {\cal D})$ for a scalar field ${\cal D}$. This model admits
rolling solutions featuring  dynamical cobordism, where the nature
of the ETW-brane
turns out to crucially depend on the value of $c$.
We will first explicitly construct negative tension, neutral  ETW-brane
solutions, showing precisely the expected scaling behavior close
to their respective cores. This is closely related to the work \cite{Raucci:2022jgw}.
Invoking the aforementioned dynamical Cobordism Conjecture
for the neutral ETW-brane 
will lead us to the upper bound 
\eq{
|c|\le c_{\rm cr}\,\qquad {\rm with}\quad   c_{\rm cr}=2 \sqrt{\frac{D-1} {D-2}}\,,
}
which is consistent with the lower bound derived from the
Trans-Planckian Censorship Conjecture \cite{Bedroya:2019snp}.

Realizing that a 0-form flux in (massive) type IIA gives  a parameter
above this bound, leads us to the explicit construction of negative
tension charged ETW-branes
in $D$-dimensions, which couple electrically to the Hodge dual
$(D-1)$-form potential. Intriguingly, these solutions exist precisely in the
complementary  region $|c|\ge c_{\rm cr}$.
For the special values $c=2\sqrt{(3D-5)/(D-2)}>c_{\rm cr}$ one
recovers the BPS solutions for  $O(D-2)$-planes, which were 
anticipated in \cite{Angius:2022aeq} as the defects closing off 
Romans-type 0-form flux backgrounds. We also provide the rules
for carrying out a  dimensional reduction from ten to $D$ dimensions
that  is consistent with the whole picture.
The previous analysis reveals  the appearance of a
lower bound $\delta\ge 2\sqrt{(D-1)/(D-2)}$
for the critical exponent in the scaling behavior
of the distance and the curvature close to the wall. 
We will argue for  a connection to the 
(sharpened) Swampland Distance Conjecture   \cite{Etheredge:2022opl} and
the (anti-) de Sitter Distance Conjecture \cite{Lust:2019zwm}.

Secondly, we also consider a generalized Blumenhagen-Font (BF) model,
which was originally introduced as the effective theory describing
the T-dual of the Sugimoto model \cite{Blumenhagen:2000dc}.
Here  the initial dilaton tadpole is already localized on a
codimension
one brane. Analogous to the special case discussed recently
\cite{Blumenhagen:2022mqw}, taking the backreaction into account one finds
again a rolling solution featuring the phenomenon of
dynamical cobordism. We explicitly construct the solution of
the codimension two ETW-brane that can close-off the initial
singular, finite size solution. In this case, we also find a critical
value of the parameter describing the exponential dilaton tadpole,
separating two  different solutions to the equations of motion, dubbed
type A and type B in the following.

\section{Generalized Dudas-Mourad model}
\label{sec_2}

Our starting point is the $D$-dimensional action of Dudas-Mourad type
in Einstein-frame
\eq{
\label{actionDMDdim}  
        S= \frac{1}{2\kappa_{D}^2}\int d^{D}x  \sqrt{-G} \left( R-\frac{1}{2} (\partial
          \phi)^2 \right) - \frac{\lambda}{\kappa_{D}^2} \int d^{D}x \sqrt{-G}\, e^{b\phi}\,,
      }
where $\phi$ denotes the dilaton\footnote{In concrete setups, $\phi$
could also be another scalar field with an exponential potential.}
  .  Here  we leave open the parameter $b$ in the exponential dilaton potential and
we assume  that the coefficient $\lambda$ is positive. 
Note that the
canonically normalized scalar is ${\cal D}=\phi/\sqrt{2}$  (in units
where $\kappa_{D}=1$).

In string theory such potentials often arise in  asymptotic regions in
dilaton field space  for
configurations with an  uncancelled dilaton tadpole.  
These could be  non-supersymmetric orientifolds with cancelled  R-R 
but uncancelled NS-NS tadpole, like the Sugimoto model \cite{Sugimoto:1999tx}. 
This was the original set-up of Dudas-Mourad \cite{Dudas:2000ff}, which
gives $D=10$, $b=3/2$ and the positive tension
$\lambda=32/(4\pi)^{\frac{1}{4}}$
of the $O9^{+}$-plane plus 32 anti D9-branes.
However, also background fluxes
can lead to such potentials, like for instance turning on the
R-R 0-form flux (Romans mass) $\tilde F$ in the 10D type IIA action. In
this case we have $D=10$, $b=5/2$ and the positive tension
$\lambda=\tilde F^2/4$. The same coefficient appears for the
ten-dimensional $SO(16)\times SO(16)$ heterotic string, where the
dilaton tadpole is generated by a non-vanishing (positive) one-loop
contribution to the cosmological constant.

\subsection{Generalized DM solution}
\label{sec:dudas}

The properties of solutions to this type of action were analyzed
in  \cite{Buratti:2021yia,Buratti:2021fiv,Angius:2022aeq}, where
it was found  that they  feature a phenomenon called dynamical
cobordism.
Solutions to the same action \eqref{actionDMDdim}, albeit with differing conventions, appeared as well in \cite{Basile:2022ypo}, in which the properties of Dudas-Mourad-like geometries upon dimensional reduction were studied. The authors also point out close connections to dynamical cobordism.
Let us discuss this in more detail.

Variation of the generalized DM action leads to the   gravity equation of motion 
\eq{
  \label{EinsteineomDM}
                  R_{MN} -\frac{1}{2} G_{MN} R -\frac{1}{2}&
                  \left(\partial_M\phi \,\partial_N \phi -\frac{1}{2} G_{MN}
                    (\partial \phi)^2\right)=-\lambda \, G_{MN} \,
                e^{b \phi} \,
                }
and the  dilaton equation of motion 
\eq{
  \label{dilatoneomDM}
              \partial_M \left( \sqrt{-G}\, G^{MN} \,\partial_N
                \phi\right)=2 b \lambda \sqrt{-G} \, e^{b \phi} \,.
}
There is no solution preserving $D$-dimensional Poincar\'e symmetry,
but one tries to solve the equations of motion for a rolling solution
preserving at least $(D-1)$-dimensional Poincar\'e symmetry.
The general ansatz for the metric is 
\eq{
  \label{metricansatzDM}
  ds^2=e^{2A(y)} ds_{D-1}^2 + e^{2 B(y)} dy^2,
}
equipped with a $y$-dependence of the dilaton  $\phi(y)$.

The resulting three a priori independent equations of motion read
\eq{
  \label{eombDM}
   (D-2) A'' +{\textstyle {(D-1)(D-2)\over 2}} (A')^2 -(D-2)A'B'
   +{\textstyle \frac 14}(\phi')^2&=-\lambda\, e^{b \phi+2B} \,\\[0.1cm]
   {\textstyle {(D-1)(D-2)\over 2}} (A')^2-{\textstyle \frac
     14}(\phi')^2&=-\lambda\, e^{b \phi+2B}\\[0.1cm]
    \phi''+((D-1) A'-B')\phi'&=2b\lambda\, e^{b \phi+2B} \,
  }
where  the prime denotes the derivative with respect to $y$.
After using the freedom to redefine the $y$-coordinate to
set $B=-{b\over 2} \phi$, the second equation motion in \eqref{eombDM}
is solved by
\eq{
         \phi'=2\epsilon_1 \sqrt{\lambda}\, \cosh f\,,\quad
         A'=\epsilon_2{\textstyle \sqrt{2\lambda\over (D-1)(D-2)} }\sinh f\,
       }
 with $\epsilon_1,\epsilon_2=\pm 1$.      
Then, the remaining two equations of motions turn out to be
related and finally lead to
\eq{
 \label{eomfinalDM} 
        f'+{(\epsilon_2 b_{\rm cr}-\epsilon_1 b)\over 2} \sqrt{\lambda} \,e^{-f}+ {(\epsilon_2 b_{\rm cr}+\epsilon_1 b)\over 2}
        \sqrt{\lambda} \,e^{f}=0\,,
}
where the (critical) parameter $b_{\rm cr}$ is defined as\footnote{Taking into
  account  the differing normalization of the dilaton, the critical parameter we
  find here is the same as the one pointed out in \cite{Basile:2022ypo}.} 
\eq{
  \label{bcrit}
  b_{\rm cr}=\sqrt{2(D-1)\over (D-2)}\,.
}
As the Ricci scalar will be of importance for the discussion of our solutions' behavior close to their boundaries, let us point out that by using the equations of motion
we obtain the  expression
\begin{equation}
\label{Ricci_DM}
R = e^{-2B(y)}\,\frac{(\phi'(y))^2}{2}+e^{b\phi(y)}\,\frac{2D}{(D-2)}\,\lambda\,.
\end{equation}

 We are looking for solutions to  \eqref{eomfinalDM} that lead to a finitely
 sized $y$-direction so that we can apply the logic of dynamical cobordism.
We recall that it  was proposed in \cite{Buratti:2021fiv,Angius:2022aeq} that (in units where $\kappa_{D}=1$ and in Einstein frame) the distance $\Delta$  and the scalar curvature $R$ scale with the distance ${\cal D}(y)$ in field space as
\begin{equation}
  \label{scalingetw}
  \Delta\sim e^{\mp{\frac{\delta}{2}} {\cal D}(y)}\,,\qquad |R|\sim e^{\pm\delta {\cal D}(y)}\
\end{equation}
with ${\cal D}(y)\to \pm\infty$ at the boundary and for a suitable
parameter $\delta>0$.

We think that a word of warning might be  appropriate
here. With the string coupling defined
as $g_s=\exp(\phi)$, where $\phi=\sqrt{2}\mathcal{D}$, the case ${\cal D}(y)\to +\infty$ is actually outside
the regime of control. Additionally, the curvature diverges at both
boundaries indicating  that we lose control over $\alpha'$-corrections. Without
reminding the reader again and again, these issues have always to be kept in mind.

In our case, the concrete solutions depend on the value
of $b$ relative to $b_{\rm cr}$. Let us discuss the region $b>b_{\rm cr}$ in detail
and then briefly provide results for the other two regions $-b_{\rm cr}<b<b_{\rm cr}$ and $b<-b_{\rm cr}$.
As we will see, the special cases  $b=\pm b_{\rm cr}$ need special treatment.

\paragraph{Region I ($b>b_{\rm cr}$):} A priori, for fixed $\epsilon_1,\epsilon_2$ there are two solution to
\eqref{eomfinalDM}, namely
\eq{
  e^{f(u)}=\epsilon_1 \sqrt{b-\epsilon_1 \epsilon_2 b_{\rm cr}\over
    b+\epsilon_1 \epsilon_2 b_{\rm cr}}\,  (\coth
  u)^{\epsilon_3} \qquad {\rm with}\ \epsilon_1 u\ge 0
 } 
with $u={1\over 2} \sqrt{\lambda( b^2-b_{\rm cr}^2)}\, y$ and $\epsilon_3=\pm 1$.
Here we have chosen an integration constant such that the solution diverges
at  $u=0$.
For the dilaton one obtains
\eq{
   \phi(u)={2\over (b-\epsilon_1 \epsilon_2 \epsilon_3 b_{\rm
       cr})}\log(\cosh u)+{2\over (b+\epsilon_1 \epsilon_2 \epsilon_3 b_{\rm cr})}\log\left|\sinh u\right|
 } 
so that  the  warp factor becomes
\eq{
  e^{B(u)}=e^{-{b\over 2} \phi}=\cosh(u)^{-{b\over b-\epsilon_1
      \epsilon_2 \epsilon_3 b_{\rm cr}}}\, |\sinh(u)|^{-{b\over
      b+\epsilon_1 \epsilon_2 \epsilon_3 b_{\rm cr}}}\,,
}
which could lead to singularities  at the two boundaries $u=0$ and
$\epsilon_1 u= \infty$.

However, only for  $\epsilon_1\epsilon_2\epsilon_3=1$ the solution turns out to
be of finite size and it only depends on the product
$\epsilon_2\epsilon_3$. Hence, after all there are only two different
solutions parametrized by the value of $\epsilon_1$.
Close to $u=0$ we can estimate 
\eq{
\begin{split}
               \Delta_{(0)} &= \frac{2}{\sqrt{\lambda(b^2-b_{\rm
                     cr}^2)}}\int_0^{u} du' e^{B(u')}\sim
               \frac{2}{\sqrt\lambda b_{cr}}
                 \sqrt{b+b_{cr}\over b-b_{cr}}\,u^{{b_{\rm
                     cr}\over b+b_{\rm cr}}}              
\end{split}
}
so that after taking into account that the canonically normalized
scalar ${\cal D}(y)$ is related to $\phi(y)$ as $\phi=\sqrt{2} {\cal D}$, one gets the scaling
\eq{
                \Delta_{(0)}\sim e^{{\sqrt{2}b_{\rm cr}\over 2} {\cal D}(y)}\,\quad {\rm
                    with} \quad {\cal D}(y)\to -\infty\  \Longrightarrow\
                  \delta_{(0)}= \sqrt{2} b_{\rm cr}\,.
}
  Close to $u=0$ we find   the following behavior of the Ricci scalar \eqref{Ricci_DM}
\eq{
R_{(0)} \sim e^{-2B(y)}\, \frac{\phi'(y)^2}{2} \sim \frac{\lambda (b-b_{\rm cr})}{2(b+b_{\rm cr})}\,u^{{-2b_{\rm cr}\over b+b_{\rm cr}}} = \frac{2}{b_{\rm cr}^2}\,\Delta_{(0)}^{-2}\,.
}
Here we kept the prefactor of the leading term, which indicates that
the curvature is positive.
Note that the scaling with the proper distance matches
the one proposed in \cite{Angius:2022aeq} so that the curvature
goes to infinity close to the boundary.

Similarly, close to the other boundary $\epsilon_1 u=\infty$, one finds the following asymptotic form of the proper distance
\eq{
                \Delta_{(\epsilon_1 \infty)}\sim  e^{-{\sqrt{2}b\over 2} {\cal D}(y)}\,\quad {\rm
                    with} \quad {\cal D}(y)\to \infty\  \Longrightarrow\
                  \delta_{(\epsilon_1 \infty)}= \sqrt{2} \, b\,.
}
While keeping track of the prefactors of both the proper distance and
$R$ itself, the Ricci scalar can be recast into the form
\eq{
\label{ricciinfty}  
R_{(\epsilon_1 \infty)} &\sim e^{-2B(y)} \, \left( \frac{2 b^2 \lambda}{b^2 - b^2_{\rm cr}} + \frac{2D}{(D-2)}\lambda \right) \\&= \frac{4(D-1)}{b^2(D-2)}\left( 1 - \frac{D}{b^2(D-2)}\right)\, \Delta_{(\epsilon_1 \infty)}^{-2}\,.
}

To close off this singular solution, we  expect the existence of
two different $(D-1)$-dimensional ETW-branes on the left and on right
of the finite interval. For $\epsilon_1=1$  we get
\eq{
  \label{closebranesa}
   {\rm ETW}^{(L,-)}_{(\delta= \sqrt{2} b_{\rm cr})}\,, \qquad    {\rm ETW}^{(R,+)}_{(\delta= \sqrt{2} b)}\,,           
 }
where the upper index indicates whether the brane is a left or a
right ETW-brane and the behavior of the dilaton at the
boundary. Note that only at the left boundary the dilaton is in the
perturbative regime $g_s\ll 1$.
The case $\epsilon_1=-1$ is supported on the negative half-space $u\le 0$ and
admits a solution that requires the left-right exchanged end-of-the-world branes
\eq{
  \label{closebranesa2}
   {\rm ETW}^{(L,+)}_{(\delta= \sqrt{2} b)}\,, \qquad    {\rm
     ETW}^{(R,-)}_{(\delta= \sqrt{2} b_{\rm cr})}  \,.         
 }

\paragraph{Region II ($-b_{\rm cr}<b<b_{\rm cr}$):} In this case, the solution is of
trigonometric type
\eq{
  e^{f(u)}=\epsilon_2\epsilon_3 \sqrt{b_{\rm cr}-\epsilon_1 \epsilon_2 b\over
   b_{\rm cr}+\epsilon_1 \epsilon_2 b}\,  (\cot
  u)^{\epsilon_3}
}
with $u={1\over 2} \sqrt{\lambda( b_{\rm cr}^2-b^2)}\, y$. As opposed to
Region I, where we encountered a singularity at $\epsilon_1 u=\infty$,
we run into a singularity already at finite $u=\epsilon_2\epsilon_3
\pi/2$.
Therefore, we restrict the domain of our solution to the interval $0 \le \epsilon_2\epsilon_3 u\le \pi/2$.
The dilaton reads
\eq{
   \phi(u)=-{2\epsilon_1 \epsilon_2 \epsilon_3\over (b_{\rm cr}-\epsilon_1 \epsilon_2 \epsilon_3 b)}\log(\cos u)+{2\epsilon_1 \epsilon_2 \epsilon_3\over (b_{\rm cr}+\epsilon_1 \epsilon_2 \epsilon_3 b)}\log\left|\sin u\right|
 }
 developing the aforementioned singularities at $u=0$ and $u=\epsilon_2\epsilon_3  \pi/2$.
Due to symmetry, we can restrict to the choice $\epsilon_2\epsilon_3=1$.
For $\epsilon_1=1$, at the  two boundaries we obtain the scalings
\eq{
                \Delta_{(0)} \sim  e^{{\sqrt{2}b_{\rm cr}\over 2} {\cal D}(y)}\quad {\rm with} \quad {\cal D}(y)\to -\infty\  \Longrightarrow\
                  \delta_{(0)}= \sqrt{2}\, b_{\rm cr}\,
}
and
\eq{
                \Delta_{(\pi/2)} \sim  e^{-{\sqrt{2}b_{\rm cr}\over 2} {\cal D}(y)}\quad {\rm with} \quad {\cal D}(y)\to \infty\  \Longrightarrow\
                  \delta_{(\pi/2)}= \sqrt{2}\, b_{\rm cr}\,.}
Analogous to Region I, the Ricci scalar at the two boundaries takes the form
\eq{
R_{(0)} \sim \frac{\lambda (b_{\rm cr}-b)}{2(b_{\rm
    cr}+b)}\,u^{-{2b_{\rm cr}\over b_{\rm cr}+b}} = \frac{2}{b_{\rm
    cr}^2}\,\Delta_{(0)}^{-2}\,,\qquad\  R_{(\pi/2)} \sim \frac{2}{b_{\rm cr}^2}\,\Delta_{(\pi/2)}^{-2}\,.
}
Hence,  at the two boundaries we find the end-of-world-branes 
\eq{
    {\rm ETW}^{(L,-)}_{(\delta= \sqrt{2}\, b_{\rm cr})}\,, \qquad     {\rm ETW}^{(R,+)}_{(\delta= \sqrt{2}\, b_{\rm cr})}\,.              
}
Again, for $\epsilon_1=-1$ one gets the left-right exchanged ETW-branes.

\paragraph{Region III ($b<-b_{\rm cr}$):} The solution is similar to the first
case
\eq{
  e^{f(u)}=-\epsilon_1 \sqrt{-b+\epsilon_1 \epsilon_2 b_{\rm cr}\over
    -b-\epsilon_1 \epsilon_2 b_{\rm cr}}\,  (\coth
  u)^{\epsilon_3}\quad {\rm with}\quad -\epsilon_1 u\ge 0
}
with  $u={1\over 2} \sqrt{\lambda( b^2-b_{\rm cr}^2)}\, y$.
For the dilaton one obtains
\eq{
   \phi(u)=-{2\over (-b-\epsilon_1 \epsilon_2 \epsilon_3 b_{\rm
       cr})}\log(\cosh u)-{2\over (-b+\epsilon_1 \epsilon_2 \epsilon_3 b_{\rm cr})}\log\left|\sinh u\right|\,.
 }
The solution is only of finite size if  $\epsilon_1 \epsilon_2
\epsilon_3 =-1$ and only depends on the product $\epsilon_2 \epsilon_3$.
Thus, for $\epsilon_1=1$,  at the  two boundaries we obtain the scalings
\eq{
                \Delta_{(0)}\sim e^{-{\sqrt{2}b_{\rm cr}\over 2} {\cal D}(y)}\,\quad {\rm
                    with} \quad {\cal D}(y)\to \infty\  \Longrightarrow\
                  \delta_{(0)}= \sqrt{2} b_{\rm cr}\,
}
and
\eq{
                \Delta_{(-\infty)}\sim e^{{\sqrt{2}|b|\over 2} {\cal D}(y)}\,\quad {\rm
                    with} \quad {\cal D}(y)\to -\infty\  \Longrightarrow\
                  \delta_{(-\infty)}= \sqrt{2} |b|\,.
}
The boundary behavior of the Ricci scalar closely resembles the one in Region I
\eq{
  R_{(0)} \sim \frac{2}{b_{\rm cr}^2}\,\Delta_{(0)}^{-2}\,,\qquad
  R_{(- \infty)} \sim \frac{4(D-1)}{b^2(D-2)}\left( 1 - \frac{D}{b^2(D-2)}\right)\, \Delta_{(- \infty)}^{-2}\,.
}
To close off this solution, we need end-of-the-world branes
\eq{
   \label{closebranesb}
   {\rm ETW}^{(L,-)}_{(\delta= -\sqrt{2} b)}\,, \qquad     {\rm
     ETW}^{(R,+)}_{(\delta= \sqrt{2} b_{\rm cr})}\,.
 }
For $\epsilon_1=-1$,  we are again led to  the left-right exchanged ETW-branes
\eq{
   \label{closebranesb2}
   {\rm ETW}^{(L,+)}_{(\delta= \sqrt{2} b_{\rm cr})}\,, \qquad     {\rm
     ETW}^{(R,-)}_{(\delta= -\sqrt{2} b)}\,.
 }

\vspace{0.2cm}

\paragraph{Special points  ($b=\pm b_{\rm cr}$):} These three regions leave out the two  points $b=\pm
b_{\rm cr}$, which indeed require a special treatment. Note that the original
Dudas-Mourad model in 10D is of this type. The solution to \eqref{eomfinalDM}
is simply
\eq{
                       {f(y)}=\mp\log\Big(\pm \epsilon\, b_{\rm cr} \sqrt{\lambda}\,
                       y\Big)\quad {\rm with}\quad \pm \epsilon y\ge 0\,,
}
where the symmetry of \eqref{eomfinalDM} allows us to choose  $\epsilon_1=\epsilon_2=\epsilon$.
This leads to
\eq{
                  \phi(y)=\pm {\lambda\over 2} b_{\rm cr} y^2 \pm{1\over b_{\rm cr}}
                  \log\Big(\pm \epsilon \sqrt{\lambda}\, y\Big)
}
and
\eq{
                         e^{B(y)}=  {1\over (\pm \epsilon\sqrt{\lambda}\, y)^{1\over 2} }  e^{-{b_{\rm cr}^2\over 4}\lambda y^2}\,.
}
At the two boundaries, we deduce the behavior
\eq{
\Delta_{(0)}&\sim e^{\pm {\sqrt{2}b_{\rm cr}\over 2} {\cal D}(y)}\,\quad {\rm with} \quad {\cal D}(y)\to \mp\infty\  \Longrightarrow\
\delta_{(0)}= \sqrt{2} b_{\rm cr}\,,\\
R_{(0)} &\sim \frac{2}{b_{\rm cr}^2}\Delta_{(0)}^{-2}
}
and
\eq{
\Delta_{(\pm \epsilon \infty)}&\sim e^{\mp{\sqrt{2}b_{\rm cr}\over 2} {\cal D}(y)}\,\quad {\rm with} \quad {\cal D}(y)\to \pm\infty\  \Longrightarrow\
\delta_{(\pm\epsilon\infty)}=\sqrt{2} b_{\rm cr}\,,\\
R_{(\pm \epsilon \infty)} &\sim \frac{2}{b_{\rm cr}^2}\Delta_{(\pm \epsilon \infty)}^{-2}\,.
}
To close off this solution, for $\pm\epsilon=1$ we need end-of-the-world branes
\eq{
   {\rm ETW}^{(L,-)}_{(\delta= \sqrt{2} b_{\rm cr})}\,, \qquad    {\rm ETW}^{(R,+)}_{(\delta= \sqrt{2} b_{\rm cr})}\,             }
 and for $\pm\epsilon=-1$ the left-right exchanged branes
\eq{
   {\rm ETW}^{(L,+)}_{(\delta= \sqrt{2} b_{\rm cr})}\,, \qquad    {\rm ETW}^{(R,-)}_{(\delta= \sqrt{2} b_{\rm cr})}\,.              } 

 \vspace{0.4cm}
Summarizing, the generalized Dudas-Mourad model features a finite size
spontaneously compactified dimension that is expected to  be closed off
at the two boundaries by the   ETW-branes shown in table \ref{tab_ETWbranes}.

\begin{table}[ht] 
  \renewcommand{\arraystretch}{1.5} 
  \begin{center} 
    \begin{tabular}{|c|c|c|c|} 
      \hline
    region &   $\epsilon_1$ &  $\epsilon_2\epsilon_3$  & ETW-branes    \\
      \hline \hline
    $b\ge b_{\rm cr}$  &  1 &  1 &  $\!\!\!{\rm ETW}^{(L,-)}_{(\delta=
                                  \sqrt{2} b_{\rm cr})}$ \quad ${\rm
                                   ETW}^{(R,+)}_{(\delta= \sqrt{2}
                                   b)}$\\
         &  -1 &  -1 &  $\!{\rm ETW}^{(L,+)}_{(\delta=
                                  \sqrt{2} b)}$ \quad\  ${\rm
                                   ETW}^{(R,-)}_{(\delta= \sqrt{2}
                                   b_{\rm cr})}$\\
      \hline
      $|b|\le b_{\rm cr}$  &  1 &  1 &  ${\rm ETW}^{(L,-)}_{(\delta=
                                  \sqrt{2} b_{\rm cr})}$ \quad ${\rm
                                   ETW}^{(R,+)}_{(\delta= \sqrt{2}
                                       b_{\rm cr})}$\\
        &  -1 &  1 &  ${\rm ETW}^{(L,+)}_{(\delta=
                                  \sqrt{2} b_{\rm cr})}$ \quad ${\rm
                                   ETW}^{(R,-)}_{(\delta= \sqrt{2}
                                     b_{\rm cr})}$\\
      \hline
      $b\le -b_{\rm cr}$  &  1 &  -1 &  ${\rm ETW}^{(L,-)}_{(\delta=
                                  -\sqrt{2} b)}$ \quad ${\rm
                                   ETW}^{(R,+)}_{(\delta= \sqrt{2}
                                      b_{\rm cr})}$\\
        &  -1 &  1 &  ${\rm ETW}^{(L,+)}_{(\delta=
                                  \sqrt{2} b_{\rm cr})}$ \quad ${\rm
                                   ETW}^{(R,-)}_{(\delta= -\sqrt{2}
                                     b)}$\\
      \hline
      \end{tabular}
          \caption{Required ETW-branes to close-off the generalized DM model.}
    \label{tab_ETWbranes} 
  \end{center} 
\end{table}

\subsection{Double dimensional reduction}

Note that in 10D the critical value is $b_{\rm cr}=3/2$, which is the
original DM model for an uncharged D9-brane/O9-plane configuration.
One would expect that via double dimensional reduction this is related to
the lower dimensional critical cases with $b=b_{\rm cr}(D)$, where we
have made the dependence on the dimension $D$ explicit.
As in \cite{Duff:1994an} (section  4.3.), this reduction involves the dilaton in a non-trivial
way via its  appearance in  the ansatz for the dimensionally reduced metric
\eq{
 \label{dimreduct} 
               G=\left(\begin{matrix} e^{\alpha \hat\phi(x)} \hat G(x) & 0
                   \\0 & e^{\beta
                     \hat\phi(x)}\,\mathbb{1}  \end{matrix}\right)\,,\qquad
               \phi=\gamma \hat\phi\,.
             }
 Here $G$ denotes the metric in 10D and $\hat G$ the one on the
 $D$-dimensional space-time. Moreover, all quantities only depend on the
 coordinates in $D$-dimensions so that 
 the Ricci scalar of $G$ becomes
 \eq{   R(G)= &-\frac{1}{2}   e^{-\alpha \hat\phi} 
                     {\textstyle \Big( 
                         {D-1\choose 2}
                       \alpha^2+(D-2)(10-D)\alpha\beta
                       +{11-D \choose 2} \beta^2\Big)}
                     (\partial\hat\phi)^2\\[0.1cm]
                     &+e^{-\alpha \hat\phi} R (\hat G)+\ldots
                   }
 where we have ignored terms that give total derivatives in the action.                  
 Plugging this ansatz into the  10D Dudas-Mourad action \eqref{actionDMDdim},
 the resulting $D$-dimensional effective action features a canonically
normalized dilaton-gravity action if
\eq{
\label{ddr1}  
  (D-2)\alpha+(10-D)\beta=0\,,\qquad
  \gamma^2=1-{4(D-2)\over (10-D)}\alpha^2\,.
}  
Now requiring that also the 10D brane action with $b=\pm{3\over 2}$ is reduced to the corresponding $D$-dimensional one
with $b(D)= \pm b_{\rm cr}$ (with $b_{\rm cr}$ defined in \eqref{bcrit}),  leads to the additional relation
\eq{
  \label{ddr2}
             b(D)=\alpha \pm {3\over 2} \gamma\,.
}
Remarkably, the three relations in \eqref{ddr1} and \eqref{ddr2}
admit  a fairly simple unique solution
\eq{
  &\alpha={(10-D)\over 4(D-2)}{1\over b(D)}\,,\qquad
  \beta=-{1\over 4\,  b(D)}\,,\qquad
   \gamma=\pm {3\over 2\,  b(D)}\\[0.2cm]
   &{\rm with}\quad 
    b(D)=\pm\sqrt{{\textstyle {2(D-1)\over (D-2)}}}\,.
}
     
\section{ETW-brane solutions for DM}
\label{ETW_DM}

In the spirit of \cite{Blumenhagen:2022mqw},
let us now investigate  whether one can construct on-shell configurations
describing  ETW-branes explicitly. These defect branes
are expected to be of codimension one in the $D$-dimensional
space-time so that in principle, they could  electrically
couple to a  $(D-1)$-form gauge potential.
Thus, one can distinguish the two cases of the  ETW-brane
being either charged or neutral.

\subsection{Neutral DM ETW-brane}
\label{sec:nonbps}

Considering  the latter case first, we are led to the ansatz for the
action of the neutral ETW-brane
\eq{
        S={1\over 2}\int d^{D}x  \sqrt{-G} \left( R-{1\over 2} (\partial
          \phi)^2 \right) - \lambda_{\rm 0} \int d^{D}x \sqrt{-g}\,
        e^{a_{\rm 0}\phi}\,
        \delta(y)\,,
      }
where $G_{MN}$ denotes the metric in $D$ dimensions  and $g_{\mu\nu}$
the metric on the $D-1$-dimensional  brane. The tension $\lambda_{\rm 0}$
and its dilaton scaling $a_{\rm 0}$ are so far unknown and need
to be determined from the consistency of the solution, if it exists at all.
The ETW-defect (to the left) is located at $y=0$,
where the non-isotropic boundary conditions are such that for $y<0$  there
is nothing, i.e. the solution is trivial and for $y>0$ there
exists a non-trivial solution to 
the resulting  gravity equation of motion 
\eq{
  \label{Einsteineom}
                  R_{MN} -{1\over 2} G_{MN} R -{1\over 2}&
                  \left(\partial_M\phi \partial_N \phi -{1\over 2} G_{MN}
                    (\partial \phi)^2\right)=\\
                  &\phantom{aaaaaaaaaaaaaaaaa}-\lambda_{\rm 0} \,\delta_M^\mu
                \delta_N^\nu \,g_{\mu\nu} \,\sqrt{g\over G}\, 
                e^{a_{\rm 0} \phi} \,\delta(y)\,.
}
In addition, there is the  dilaton equation of motion 
\eq{
  \label{dilatoneom}
              \partial_M \left( \sqrt{-G}\, G^{MN} \,\partial_N
                \phi\right)=2 a_{\rm 0} \lambda_{\rm 0} \sqrt{-g} \,
              e^{a_{\rm 0} \phi} \,\delta(y)\,.
}
The question is whether there exists a choice of the tension
$\lambda_{\rm 0}$ and the coefficient $a_{\rm 0}$ such that an ETW solution
exists that reproduces the general scaling behavior of \cite{Angius:2022aeq}
close to the defects. 

The general ansatz for the metric and the dilaton is 
\eq{
  \label{metricansatz1}
  ds^2=e^{2{A}(y)} ds_{D-1}^2 + dy^2\,,\qquad \phi=\phi(y)\,,
}
which still preserves the longitudinal $(D-1)$-dimensional Poincar\'e
symmetry and where we have used the freedom of a coordinate transformation
to set $g_{yy}=1$. After a few steps, the three resulting equations of motion can be brought to
the form
\eq{
\label{eomDef}
           &\phi'=\pm \sqrt{2(D-1)(D-2)}\,A' \\[0.1cm]
           &  A'' + (D-1) (A')^2=- {\textstyle {1\over (D-2)}}\, \lambda_{\rm 0}
           \,e^{a_{\rm 0}\phi}\,\delta(y)\\[0.1cm]
             &  A'' + (D-1) (A')^2=\pm {\textstyle {\sqrt{2\over
                 (D-1)(D-2)}}}\, a_{\rm 0} \,\lambda_{\rm 0} \,e^{a_{\rm 0}\phi}\,\delta(y)\,.
} 
Thus, consistency of the last two equations implies that there can
only exist a solution for
\eq{
  a_{\rm 0}=\mp \sqrt{{(D-1)\over 2(D-2)}}=\mp b_{\rm cr}/2\,.
}

Solving the equations of motion \eqref{eomDef} in the bulk we find                     
\eq{
\label{sol_A}
 A(y) = \frac{1}{D-1} \log\Big\lvert (D-1)y-c_1 \Big\rvert\,,
 }      
where  we have set an integration constant to zero by a redefinition
of the $x$-coordinates.
Then, from the other independent equation of motion we immediately get
 \eq{
 \label{sol_phi}
 \phi(y) = \phi_0 \pm \sqrt{\frac{2(D-2)}{D-1}} \log\Big|(D-1)y-c_1\Big|\;.
}
Notably, close to all boundaries characterized by $\delta = \sqrt{2}\,
b_{\rm cr}$, both \eqref{sol_A} and \eqref{sol_phi} can be
exactly matched with $A(\Delta)$ and $\phi(\Delta)$ of the singular
DM like solutions from section \ref{sec_2}
once the latter are expressed in terms of the corresponding
proper distance $\Delta$. We emphasize
that this entails that the boundary conditions of the
neutral ETW-branes, which we will impose next, can indeed be consistently
implemented in the original DM like solutions.

Imposing now the boundary condition at $y=0$ means that the first derivative
of the warp factor $A(y)$ (and $\phi(y)$ accordingly) jumps like
  \eq{
    \Delta A'\Big\vert_{y=0} 
    = -{1\over (D-2)} \,\lambda_{\rm 0}\,  e^{a_{\rm 0} \phi(0)} \,,
 }
where 
\eq{
  A'(y) =
    \begin{cases}
      1 \over (D-1)y-c_1 & y>0\\
      0 & y \le 0\\
    \end{cases}       
}
imposes the condition that for negative $y$ there is ``nothing''.
This leads to the relation
\eq{
\label{boundicon}  
       -{1\over c_1}=-{\lambda_{\rm 0}\over (D-2)} \,  e^{a_{\rm 0}
         \phi_0} {1\over |c_1|} \,.
     }
This determines the integration constant $\phi_0$ in terms of the
ETW-brane tension $\lambda_{\rm 0}$
 \eq{
   \lambda_{\rm 0}={\rm sgn}(c_1) (D-2) \, e^{-a_{\rm 0} \phi_0} \,.
 }
For $c_1$ positive, we would have a singularity in the $y$-positive
region, namely at $y=c_1/(D-1)$. Therefore, one could conclude that
we need to choose $c_1$ negative to have a physically reasonable
solution. This implies that the tension of the ETW-brane is negative.

For finite  value of $c_1$ we find the following behavior of the warp factor
and the dilaton when approaching  $y=0$ from above
 \eq{
   \lim_{y\to 0^+}  A(y)&={1\over (D-1)} \log|c_1|\,,\\[0.1cm]
    \lim_{y\to 0^+}  \phi(y)&=\phi_0\pm  {\textstyle \sqrt{\frac{2(D-2)}{D-1}}}   \log|c_1|\,.
} 
Imposing that for $y<0$ we have nothing means that the metric there is
vanishing, i.e. the warp factor $A$ should go to minus infinity. 
This means that eventually, we should take the limit $c_1\to 0^-$.

Thus, we have found two solutions  corresponding to  neutral ETW-branes 
of codimension one carrying  negative tension and featuring a brane action
\eq{
   S_{\rm ETW}=  -\lambda_{\rm 0} \int d^{D-1}x\,  \sqrt{-g} \, \exp\left( \mp
     {\textstyle \sqrt{{(D-1)\over 2(D-2)}}}\phi(y)\right) \,.
 }
Then we find $\lim_{y\to 0^+}  \phi(y)=\mp \infty$.

Since there is no warp factor in front of the $dy^2$-term in \eqref{metricansatz1},
the distance to the defect brane at $y=0$ is simply given
by
\eq{
             \Delta_{(0)}=y\sim \exp\left(\pm {\textstyle
                 \sqrt{(D-1)\over (D-2)}} {\cal D}(y)\right)
}
from which we can read off
\eq{
\delta_{0} = 2 \sqrt{\frac{D-1} {D-2}}\,.
}
Analogously to the DM solution, one can straightforwardly simplify the
expression for the Ricci scalar with the equations of motion arriving
for $y\ne 0$ at
\begin{equation}
  \label{Ricci_nETW}
R \sim \frac{\phi'(y)^2}{2} = \frac{(D-2)}{(D-1)}\, y^{-2} = \frac{2}{b_{\rm cr}^2}\, \Delta_{(0)}^{-2}\,.
\end{equation}
Notably, not only does the scaling behavior of the Ricci scalar in terms of the proper distance agree with the Ricci scalar at the boundaries of the Dudas-Mourad-like model,
but we also  get the same prefactor $2/b_{\rm cr}^2$ as in the case of the $\delta= \sqrt{2} b_{\rm cr}$ boundaries.

Hence, the neutral brane with $a_0=\mp b_{\rm cr}/2$ has precisely the properties
of the ${\rm ETW}^{(L,\mp)}_{(\delta= \sqrt{2} b_{\rm cr})}$ brane.
Completely analogous, one can construct solutions corresponding to  ETW-branes to
the right. Using the same solutions for $A(y)$ and $\phi(y)$ in the
bulk,
the only change is that the relation \eqref{boundicon} acquires an
extra minus sign
\eq{
       {1\over c_1}=-{\lambda_{\rm 0}\over (D-2)} \,  e^{a_{\rm 0}
         \phi_0} {1\over |c_1|} \,
     }
     so that $c_1>0$ with the tension still coming out negative.
The solution with $a_0=\pm b_{\rm cr}/2$ has precisely the properties
of the ${\rm ETW}^{(R,\pm)}_{(\delta= \sqrt{2} b_{\rm cr})}$ brane.

\subsubsection*{Double dimensional reduction}

Let us check how this ETW-action behaves under the dimensional
reduction discussed for the DM solution.
Requiring that also the 9D ETW-brane action with $a_0=\pm{3\over 4}$
is reduced to the corresponding $(D-1)$-dimensional one
with $a_0(D)=\pm b_{\rm cr}(D)/2$,  leads to the condition
\eq{
  \label{ddr4}
  a_0(D)={\alpha\over 2}\pm {3\over 4}\gamma\,.
}
Hence, we get precisely the same condition as in \eqref{ddr2}
and consistently, via dimensional reduction both the
original DM action and the neutral ETW-brane action
behave as expected.

\subsection{Dynamical Cobordism Conjecture: DM neutral case}

First, let us observe that in ten dimensions  one gets $a_{0}=\mp
{3\over 4}$ so that  the string frame brane action becomes
\eq{
  S\left({{\rm ETW}^{(L(R),\mp)}}\right)=  - T \int d^{10}x \sqrt{-g}\, \delta(y)
                         \begin{cases} e^{-3\phi}\\ e^{-{3\over
                               2}\phi}\,.\end{cases}
}
This  does not correspond to any known brane tension so that we
conclude that these neutral 9-dimensional ETW-branes are new
objects.

As mentioned these four non-isotropic  solutions have the right properties to be identified
with the ETW-branes needed to close off the singular Dudas-Mourad
type solutions from the previous section, at least at those boundaries
with $\delta=\sqrt{2} b_{\rm cr}$.
Thus, applying now the dynamical version  of the Cobordism Conjecture,
we arrive at a swampland conjecture of the form:
\begin{quotation}
\noindent{\it The neutral ETW-brane can completely close off the singular DM solution
only for the initial parameter lying  in the interval $|b|\le b_{\rm cr}$.}
\end{quotation}

\noindent As we have seen, outside this
region one would need a different  ETW-brane with $\delta\ne \sqrt{2}
b_{\rm cr}$ at one of the two boundaries.

Let us express the results in terms of the coefficient
 $c=\sqrt{2} b$ appearing in the exponential $\exp(c {\cal D})$ for the
 canonically normalized field ${\cal D}(y)$.
Combining the former bound with the one  derived from the Trans-Planckian
Censorship Conjecture (TCC) \cite{Bedroya:2019snp} we get that the values of $c>0$
consistent with the Dynamical Cobordism Conjecture
are restricted to the range
\eq{
 \label{cobordtccconstr} 
                     {2\over \sqrt{(D-1)(D-2)} }\le c\le  2 \sqrt{\frac{D-1} {D-2}}\,.
 }
Note that the lower TCC bound is indeed smaller than the upper bound
for $D\ge 3$, where for $D=2$ their ratio is finite and exactly one.

Note that this bound still  allows values  of $c$ larger than the
upper bound, but only at the expense of finding explicit
ETW-brane solutions that can close off the DM solution at the second
boundary. For getting an idea what we should look at, consider
the example of the 10D type IIA action with a non-vanishing 0-form
flux. As mentioned at the beginning of section \ref{sec_2}, in this case we
have the DM action \eqref{actionDMDdim} with $D=10$ and $b=5/2$. Hence, we have
$b>b_{\rm cr}=3/2$.   Does this mean that this solution cannot be
closed off? As mentioned already in \cite{Buratti:2021fiv}, in this case one expects
the ETW-brane to also carry magnetic   charge under the 0-form field
strength. This makes stacks of $O$8/$D$8-branes natural candidates.

\subsection{Charged DM ETW-brane solution}

Let us now analyze whether one can construct on-shell configurations
describing such charged  ETW-branes explicitly. Since these defect branes
are of codimension one they can indeed couple electrically to a
$(D-1)$-form gauge potential $C$, whose field strength is Hodge-dual
to the 0-form $\tilde F$ generating the dilaton tadpole in the DM action.
Hence, for such a brane we are led to consider  the action
\eq{
\label{chargedaction}  
        S=&{1\over 2}\int d^{D}x  \sqrt{-G} \left( R-{1\over 2} (\partial
          \phi)^2 -{1\over 2}  \, e^{b_c\phi} \,|F|^2\right) \\
         &- \lambda_c  \int d^{D}x \sqrt{-g}\,
        e^{a_c\phi}\,
        \delta(y) + \mu_c\int_\Sigma  C
      }
 with the field strength $F_{M_1\ldots M_{D}}=D\, \partial_{[M_1}
 C_{M_2\ldots M_D]}$  and
 \eq{
 \label{defkineticflux}  
   |F|^2= {1\over D!}\, G^{M_1 N_1} \ldots G^{M_D N_D}\, F_{M_1\ldots M_{D}}
   F_{N_1\ldots N_{D}}\,.
 }
We initially consider  the coefficients $b_c\,,a_c\,$, the brane tension $\lambda_c$ and its charge $\mu_c$ as
 free parameters and will determine them  from the consistency of the solution.

The general ansatz for the metric, the dilaton and the $(D-1)$ form is 
\eq{
  \label{metricansatz2}
  ds^2=e^{2{A}(y)} ds_{D-1}^2 + dy^2\,,\qquad \phi=\phi(y)\,,\qquad
  C_{01\ldots D-2}=C(y)\,,
}
which still preserves the longitudinal $(D-1)$-dimensional Poincar\'e
symmetry. 
The equations of motion for the metric components $g^{yy}$ and
$g^{\mu\nu}$ read
\eq{
   {\textstyle {(D-1)(D-2)\over 2}} (A')^2 -{\textstyle
         \frac 14}(\phi')^2&+{\textstyle \frac 14} e^{-2(D-1)A +b_c\phi} (C')^2=0\,,\\[0.3cm]
        (D-2) A'' +{\textstyle {(D-1)(D-2)\over 2}} (A')^2 &+{\textstyle
         \frac 14}(\phi')^2\\
         &+{\textstyle \frac 14} e^{-2(D-1)A +b_c\phi} (C')^2=-\lambda_c\, e^{a_c \phi} \, \delta(y)\,.       
}  
The resulting dilaton and $(D-1)$-form equation of motion take the form
\eq{
  \phi'' +(D-1) A' \phi' +{\textstyle {b_c\over 2}}\, e^{-2(D-1)A +b_c\phi} (C')^2&=2a_c\lambda_c\, e^{a_c \phi} \, \delta(y)\,,\\[0.3cm]
   C'' +b_c  \phi' C' -(D-1) A' C' &=2\mu_c\, e^{(D-1)A -b_c\phi} \, \delta(y)\,.
     }
Making the ansatz $\phi=2\beta A+\phi_0\,$, from the metric and dilaton
equations of motion one can determine $\beta={b_c\over 2}(D-2)$
being left with  an equation for $A(y)$
\eq{
  \label{finaleqfora}
                   A'' +{b_c^2\over 2}(D-2) (A')^2= -{\lambda_c\over (D-2)}
                   e^{a_c\phi} \delta(y)\,.
}
In the bulk this features the solution
\eq{
  \label{abulksoli}
       A(y) = \frac{2}{b_c^2 (D-2)} \log\left\vert {b_c^2\over 2}\,(D-2)\,y-c_1 \right\vert\,.
     }
Note that $b_c$ is still a free parameter.
The final equation fixing $C(y)$ only admits a real solution for
$b_c^2\ge b^2_{\rm  cr}$ (with $b_{\rm  cr}$ defined in \eqref{bcrit})
and determines
\eq{
 \label{finalCy} 
       C(y)=\pm{2\over \sqrt{b_c^2-b_{\rm cr}^2}} { e^{-{b_c\over 2}\phi_0}\over
         \Big({b_c^2\over 2}\, (D-2)\, y -c_1 \Big)^{1-{(b^2_{\rm cr}/ b_c^2)}}}\,.
     }
Consistency of the boundary conditions then directly fixes the
coefficients in the brane action as
\eq{
  \label{finalvaluespara}
  a_c=-{b_c\over 2}\,,\qquad\mu_c=\pm {1\over 2}\sqrt{b_c^2-b_{\rm
      cr}^2} \,\lambda_c\,.
}
Hence, for the critical values $b_c=\pm b_{\rm cr}$ we get a neutral
ETW-brane solution, which nicely connects to the neutral solution
presented in section \ref{sec:nonbps}. Moreover, for
\eq{
  b_{c,{\rm BPS}}=\pm\sqrt{4+b_{\rm cr}^2}=\pm\sqrt{2(3D-5)\over (D-2)}
}  
we get the BPS case with $\lambda_c=|\mu_c|$. 

Finally, one needs to implement the boundary conditions at $y=0$
in the equation \eqref{finaleqfora}.
This  is analogous to the final equation for the uncharged ETW-brane.                 
The bulk solution \eqref{abulksoli} is leading to the dilaton            
  \eq{
 \label{sol_phi_ch}
 \phi(y) = \phi_0 + {2\over b_c} \log\left\vert {b_c^2\over 2}(D-2)y-c_1 \right\vert\;.
}
As in the neutral case, close to the $\delta = \sqrt{2}\,b$ boundaries
both \eqref{abulksoli} and \eqref{sol_phi_ch} are reproduced by
expressing the DM  solutions
in terms of the appropriate proper distance and using the sign-flip
\eqref{signflip}.

Proceeding analogously to  the uncharged case, imposing now the
left/right boundary conditions
at $y=0$ leads to
\eq{
  \label{tension_charged}
  \lambda_c=-(D-2) \, e^{{b_c\over 2}\phi_0} \,,\qquad \mu_c=\mp
    {(D-2)\over 2}\sqrt{b_c^2-b_{\rm
      cr}^2} \,e^{{b_c\over 2}\phi_0}\,.
 }
Again avoiding  a singularity in the $y$-positive/negative region, 
forced us  to choose $c_1$ negative/positive with finally taking the limit $c_1\to 0^\mp$.
This leads  to an ETW-brane with negative tension 
$\lambda_c$. 
For the distance to the brane we get
\eq{
             \Delta_{(0)}=y\sim \exp\left(\pm {\textstyle
                 {\sqrt{2}\, b_c\over 2}}\, {\cal D}(y)\right)
}
from which we can read off $\delta_{c}=\sqrt{2} b_c$. 
Due to the appearance of a dynamical gauge potential the Ricci scalar 
  cannot be simplified to $\phi'(y)^2/2$ like for the neutral brane
  \eqref{Ricci_nETW}.
  Instead we obtain
\eq{
R &\sim \frac{(D-1)}{(D-2)}\left(1-\frac{D}{b_{\rm
      c}^2\,(D-2)}\right)\phi'(y)^2\\[0.1cm]
 &\sim \frac{4(D-1)}{b_{\rm c}^2\,(D-2)}\left(1-\frac{D}{b_{\rm c}^2\,(D-2)}\right)\Delta_{(0)}^{-2}\,,
}
which agrees with the corresponding boundary behavior \eqref{ricciinfty} of the
supercritical DM like model.

Before we conclude let us   check that indeed this charged ETW-brane solution
carries  a
constant magnetic dual zero-form flux (Romans mass) $\tilde F$.
There is one subtlety, which is related to the fact that the kinetic
term for $\tilde F$ will have a different dilaton term in front.
We only get the wanted result if this term scales as
\eq{
                    S=&{1\over 2}\int d^{D}x  \sqrt{-G} \left(
                      -{1\over 2}  \, e^{-b_c\phi} \,|\tilde F|^2\right)\,, 
}
which means that the coefficient $b=b_{\rm DM}$ in the DM action
is related to the one in the charged
ETW-brane action  as
\eq{
  \label{signflip}
                          b_c=-b_{\rm DM}\,.
}
As a consequence, the correct duality transformation is
\eq{
            \tilde F=e^{b_c\phi} \,\star F=-e^{b_c\phi} {1\over
              \sqrt{-G}} \,\partial_y C 
}
where in the last step we have used the  explicit solution.
This can be straightforwardly evaluated yielding
\eq{
  \label{0formflux}
              \tilde F=- (D-2) \sqrt{b_c^2-b_{\rm cr}^2}\, e^{{b_c\over
                  2}\phi_0}=\pm 2\mu_c\,.
 }
Hence, one observes that the magnetic  dual 0-form flux is indeed constant
and related to the charge of the ETW-brane. In particular, flipping
the sign of the 0-form flux can be compensated by flipping
the sign in the solution \eqref{finalCy} for the $(D-1)$ form $C(y)$.
Invoking that a
non-vanishing 0-form flux in the generalized DM action \eqref{actionDMDdim} generates a
$\lambda=|\tilde F|^2/4$ we can also write the relation \eqref{0formflux}
as $\lambda=\mu_c^2$.

Taking the relation \eqref{signflip} into account, we can now conclude that
we have found charged ETW-brane solutions that carry all the features
expected for the left and right end-of-world branes:
\eq{
      b_{\rm DM}>b_{\rm cr}:& \quad   {\rm ETW}^{(L/R,+)}_{(\delta= \sqrt{2}\,
        b_{\rm DM})}\\[0.1cm]
      b_{\rm DM}<-b_{\rm cr}:& \quad   {\rm ETW}^{(L/R,-)}_{(\delta= -\sqrt{2}\,
        b_{\rm DM})}\,.
    }
Note the correlation between the dilaton behavior and the sign of
$b_{\rm DM}$,
which is just right to close off the corresponding DM solutions listed
in table \ref{tab_ETWbranes}.

\subsubsection*{Double dimensional reduction}

Let us show that the action \eqref{chargedaction} and its charged
ETW-brane solution is consistent upon
double dimensional reduction. For this purpose the computation
from the end of section \ref{sec:dudas} needs to be generalized to the charged case,
which admits a free parameter $b_c$.
Of course, one should recover the relations \eqref{dimreduct}
for the special  case of a neutral ETW-brane with $b_c^2=b_{\rm
  cr}^2$.

The starting point is the 10D action \eqref{chargedaction} with a coefficient
$b_c=b_c(10)$ and the two relations in \eqref{finalvaluespara}
satisfied. Here we consider the branch that includes the BPS brane,
i.e. the positive sign in \eqref{finalCy} and \eqref{finalvaluespara}.
The dilaton gravity part of the computation proceeds as in the neutral
case so that we arrive at the two relations in \eqref{ddr1}.

Consistency of the ETW-brane action now requires that upon dimensional
reduction one needs to satisfy
\eq{
  \sqrt{b^2_c(D)-b^2_{\rm cr}(D)}= \sqrt{b^2_c(10)-b^2_{\rm cr}(10)}
}
with $b_{\rm cr}(10)=3/2$ and the resulting dimensionally reduced coefficient
\eq{
             b_c(D)=-\alpha +\gamma\, b_c\,
}
in $D$ dimensions. Thus, we have three relations for the three
coefficients $\alpha,\beta,\gamma$, which admit  a unique solution
\eq{
  &\alpha=-{(10-D)\over 4(D-2)}{1\over b_c(D)}\,,\qquad
  \beta={1\over 4\,  b_c(D)}\,,\qquad
   \gamma={b_c\over  b_c(D)}\,,\\[0.2cm]
   &{\rm where}\quad 
    b_c(D)=\pm\sqrt{b_c^2+{\textstyle {(10-D)\over 4(D-2)}}}
}
with the plus sign in the  region  $b_c\ge 3/2$ and the negative sign
in the region $b_c\le -3/2$.

It is straightforward to confirm that  for $b_c=\pm 3/2$ one gets
$b_c(D)=\pm b_{\rm cr}(D)$. Moreover, in the BPS case $b_c=\pm 5/2$
one finds
\eq{
               b_c(D)=b_{c,{\rm BPS}}=\pm \sqrt{2(3D-5)\over (D-2)} \,     
}
so that dimensional reduction maps 10D  BPS ETW-branes to the lower
dimensional BPS ETW-branes.

\subsection{Dynamical Cobordism Conjecture: DM charged case}
\label{DCC_charge}

We have seen in the previous section that if the exponential in the
Dudas-Mourad action \eqref{actionDMDdim} carries a coefficient
$|b|\ge b_{\rm cr}$, 
then the corresponding finite size solution
can be closed off  by a charged codimension one
ETW-brane that is electrically charged under the dual ($D-1)$-form.
More concretely, on one boundary the finite interval can be closed off by an uncharged
ETW-brane (with $\delta=\sqrt 2  b_{\rm cr}$) and the other boundary by
a  charged one with  (with $\delta=\sqrt 2 b$).
Like in the neutral case, applying now the dynamical version  of the Cobordism Conjecture,
we arrive at a swampland conjecture of the form:
\begin{quotation}
\noindent{\it The charged ETW-brane can  close off the singular DM solution
only for the initial parameter lying  in the interval $|b|\ge b_{\rm
  cr}$.  For both the neutral and the charged case, the
scaling parameter is bounded from below as $\delta\ge \delta_{\rm cr}$ with
\eq{\delta_{\rm cr} =   2 \sqrt{(D-1)\over (D-2)}\,.}
 }
\end{quotation}

\vspace{0.1cm}
\noindent
Hence, we have identified all (on-shell) ETW-branes that can close off the
DM solutions of section \ref{sec_2} for
\eq{
  & |b| \le \sqrt{2(D-1)\over (D-2)}\,,\quad {\rm 2\ neutral\ ETW}\\[0.2cm]
  & |b| \ge \sqrt{2(D-1)\over (D-2)}\,,\quad {\rm 1\ neutral\ +\ 1\ charged \ ETW}\,.
}

The second regime contains the BPS case. 
For  $D=10$ the massive type IIA case has a dilaton tadpole with
$b_{\rm DM}=5/2$ leading to the parameters $b_c=-5/2$ and $a_c=5/4$
in the charged ETW-brane action. In string frame the brane action
features a coefficient $a_c^{\rm st}=-1$, which is precisely
the action of an  $O8$-plane\footnote{One expects that the $O8$-planes provide
the defects to break an initial cobordism group $\Omega^\xi_0=\mathbb
Z$. It would be interesting to determine what the actually relevant
structure $\xi$ is. Furthermore, this implies an interesting interplay between geometric and topological statements, which warrants further investigation.} (carrying negative tension).
Hence we conclude that  in the BPS case with  $b_{\rm DM}=\sqrt{2(3D-5)\over (D-2)}$,
the charged ETW-branes can be considered
as $O(D-2)$ planes.

Let us make two comments about the relation of our bound to existing
results in the literature.
First, our lower bound on $\delta$ is in agreement with
\cite{Angius:2022aeq}, where the authors looked at general exponential
dilaton potentials for ETW-branes. In their analysis a positive
potential $V(\phi)>0$ always implies $\delta >  \delta _{\rm cr}$,
whereas the critical case  $\delta =  \delta _{\rm cr}$ is somewhat special as it corresponds to a negligible potential compared to the kinetic term $\phi'^2$.

Second, by comparing to the investigation of exponential scalar
potentials in the context of holography \cite{Bedroya:2022tbh}, the
latter charged ETW-brane regime of \ref{actionDMDdim} is matched with
a regime, where the  contribution of the kinetic term to the Hubble
energy is of the same order as the contribution from the potential.
The neutral ETW-brane regime on the other hand corresponds to a
dominant kinetic term contribution to the Hubble energy.

\subsubsection*{A comment on the $SO(16)\times SO(16)$ heterotic string}

As mentioned at the beginning of section \ref{sec_2}, the
non-supersymmetric, tachyon-free $SO(16)\times SO(16)$ heterotic
string has a one-loop dilaton tadpole, which also gives
rise to a DM type action with a coefficient $b=5/2$. As we have
seen, one of the required nine-dimensional ETW-branes carries charge and couples electrically
to a 9-form gauge potential. However, in contrast to the massive type
IIA case, where the tadpole was generated by a tree-level 0-form flux,
here we have no reason to expect that the ETW-brane carries any
charge.
Hence, before claiming that the $SO(16)\times SO(16)$ heterotic
string contains a new yet undiscovered charged  $O8$-brane
(of negative tension), we should
consider alternative resolutions of this issue.

As mentioned at the beginning, we can actually only trust the
analysis for an ETW-brane where the string coupling goes to zero close to its
core. These would be all the ${\rm ETW}^{(L/R,-)}_\delta$ branes.
For the other half of branes, namely  ${\rm ETW}^{(L/R,+)}_\delta$,
there
might exist an S-dual frame where the analysis might be carried out in
a reliable way, but that is often not straightforward to establish.
Now, for $b=5/2$ we need on one side the neutral ETW-branes
\eq{
                {\rm ETW}^{(L/R,-)}_{\delta=\sqrt{2} b_{\rm cr}
                }\,,\qquad b_{\rm cr}=3/2\,,
}
which are always at weak string coupling. These are just solutions of
the universal dilaton-gravity sector and as such can persist for the  $SO(16)\times SO(16)$ heterotic
string, as well.
However, closing-off the other boundary requires charged ETW-branes
\eq{
                {\rm ETW}^{(L/R,+)}_{\delta=\sqrt{2} b }\,,\qquad
                b=5/2\,,
}                
which are always at strong string coupling, where the loop-expansion
of the vacuum energy cannot be trusted. Hence, the correlation between
the sign of $b$ and the dilaton behavior, prevents us from 
automatically concluding that the $SO(16)\times SO(16)$ heterotic
string contains a new charged $O8$-plane.

\subsubsection*{Relation to Swampland Distance Conjectures}

As already pointed out in \cite{Buratti:2021fiv}, the fact that
$\phi(y)\to \pm \infty$ as we approach one of the two singularities
  at $y=0$ or $y= \infty$ implies an infinite distance limit in the
  dilaton moduli space. Therefore,  due to  the Swampland Distance
  Conjecture \cite{Ooguri:2006in} we expect the appearance of a tower
  of exponentially light  states
\eq{
m \sim e^{\mp\lambda_{\rm SDC} {\cal D}}\,,
}
where $\lambda_{\rm SDC}$ is an $O(1)$ constant and ${\cal D}$ is the distance in moduli space. Recently, in \cite{Etheredge:2022opl} the authors made a strong effort to sharpen the conjecture and proposed constraints on $\lambda_{\rm SDC}$. In the context of our analysis the upper bound 
\eq{
  \label{sdcupper}
\lambda_{\rm SDC} \leq \sqrt{(D-1)\over (D-2)}\,,
}
based on the Emergent String Conjecture \cite{Lee:2019xtm,Lee:2019wij}, is of special interest.

We have seen that close to the boundary, the positive Ricci-scalar
goes to infinity as
\eq{
R\sim e^{\pm\delta {\cal D}(y)}\,.
} 
Combining this with the mass scaling of the Swampland Distance Conjecture
we obtain the scaling relation 
\eq{
  m_{\rm SDC} \sim R^{-\alpha}\qquad{\rm with}\quad
  \alpha =  \frac{\lambda_{\rm SDC}}{\delta}\,.
}
As already highlighted at the beginning of this section \ref{DCC_charge},  $\delta$  differs from $\sqrt{2} |b_{\rm cr}|$ only in the parametric regime $|b| > |b_{\rm cr}|$, where it becomes $\sqrt{2} |b|$.
Consequently, this places a lower bound on $\delta$
\eq{
\delta \geq \sqrt{2} b_{\rm cr} = 2 \sqrt{(D-1)\over (D-2)}\,,
}
which together with the upper bound \eqref{sdcupper} on $\lambda_{\rm SDC}$ implies
the  upper bound
\eq{
\alpha\leq \frac{1}{2}\,.
}
The scaling and the bound on $\alpha$ is eerily similar to the (Anti-) de Sitter Distance Conjecture \cite{Lust:2019zwm}, which proposed that in the limit of an asymptotically vanishing cosmological constant $\Lambda$ a tower of massive states should become light 
\eq{
m_{\rm ADC} \sim |\Lambda|^{\beta}\,.
}
In the case of de Sitter space, $\beta$ is bounded from above by $\frac{1}{2}$, as well\footnote{This stems from the Higuchi bound \cite{Higuchi:1986py}.}. However, we observe two key differences. First, our warped space-time solution to the DM like model does not become de Sitter close to the boundaries, it is merely positively curved.
Second, in contrast to  the (Anti-) de Sitter Distance Conjecture,
we do not have vanishing curvature close to the boundary but instead the Ricci scalar is diverging.
Of course, this poses the question, whether the similarity is entirely coincidental or whether it points towards a sensible generalization of the (Anti-) de Sitter Distance Conjecture 
to space-times, which are not strictly (anti-) de Sitter and/or with asymptotically infinite curvature. We leave this as an open question for future work.

\section{ETW-brane for generalized BF-model}
\label{sec:ETW-BF}

So far we  discussed the generalized Dudas-Mourad model, which
has revealed an intricate structure of running dilaton solutions and their
end-of-the-word defects. Already in the year 2000, in
\cite{Blumenhagen:2000dc} the DM model has been generalized to a
codimension one dilaton tadpole, as it would occur in the T-dual
Sugimoto model or for a non-BPS $D8$-brane in type I string theory. 
A non-isotropic ETW-brane solution for a special case of this  model was recently
presented in \cite{Blumenhagen:2022mqw}. In this section we generalize this solution to
arbitrary dimension and coefficient $a$ in the running dilaton potential.

\subsection{Generalized BF-solution}
\label{sec:genBFsol}

We consider a  neutral domain wall  configuration in ten dimensions carrying positive tension  and being located at the position $r=0$ in the transversal directions. 
Since it carries no other gauge charge, at leading order its supergravity
action assumes the form
\eq{
  \label{actionBFgen}
        S={\frac{1}{2}}\int d^{D}x  \sqrt{-G} \left( R-{\frac 12} (\partial \phi)^2 \right) - \lambda \int d^{D}x \sqrt{-g}\, e^{a \phi}\, \delta(r)\,.
      }
The resulting  equations of motion are the same as in \eqref{Einsteineom}
and \eqref{dilatoneom}.

In general, these equations do not admit a solution preserving $(D-1)$-dimensional Poincar\'e invariance. 
However, in \cite{Blumenhagen:2000dc} a solution was found that preserved $(D-2)$ Poincar\'e invariance featuring a single  non-trivial longitudinal direction $y$.
The general ansatz for the metric was 
\eq{
  \label{metricansatz3}
  ds^2=e^{2{\cal A}(r,y)}\, ds_{D-2}^2 + e^{2{\cal B}(r,y)}\,( dr^2 +dy^2)
}
with a separated dependence of the warp factors $\mathcal{A}$, $\mathcal{B}$ and the dilaton $\phi$ on the coordinates $r$
and $y$, i.e. 
\eq{
       &{\cal A}(r,y)=A(r) + U(y)\,,\quad {\cal B}(r,y)=B(r) +
       V(y)\,,\\
       &\phi(r,y)=\chi(r) + \psi(y)\,.
}
For such a separation of variables, by redefining $r$ and $y$ the
ansatz \eqref{metricansatz3} is actually the most general one.

The equations of motion lead to five a priori independent equations.
The one related to the variation  $\delta G^{\mu\nu}$ is
\begin{eqnarray}
 \label{eommunub}
  &&\phantom{a}\bigg( (D-3) A'' +{(D-2)(D-3)\over 2} (A')^2 +B'' +{\textstyle \frac 14}(\chi')^2\bigg)\\
  &&+\bigg( (D-3) \ddot U  +{(D-2)(D-3)\over 2} (\dot U)^2+ \ddot V + {\textstyle \frac 14} (\dot
     \psi)^2\bigg) =-\lambda\, e^{B+V}\, e^{a \phi} \, \delta(r)\,.\nonumber
\end{eqnarray}     
 The prime denotes the derivative with respect to $r$ and the dot the derivative with respect to $y$.
 For the variation $\delta G^{rr}$, we obtain
\eq{
  \label{eomrr}
   &\bigg( {(D-2)(D-3)\over 2} (A')^2 +(D-2) A' B' - {\textstyle{\frac
       14}}(\chi')^2\bigg) \\
   +&\bigg( (D-2) \ddot U+ {(D-1)(D-2)\over 2} (\dot U)^2- (D-2) \dot U\dot V + {\textstyle{\frac 14}} (\dot
  \psi)^2\bigg)=0,\\[0.2cm]
}
and for  variation $\delta G^{yy}$
\eq{
  \label{eomyy}
  &\bigg( (D-2) A'' + {(D-1)(D-2)\over 2} (A')^2 -(D-2)A' B'
  +{\textstyle \frac 14}(\chi')^2\bigg)\\
  +&\bigg(  {(D-2)(D-3)\over 2} (\dot U)^2+ (D-2) \dot U\dot V - {\textstyle \frac 14} (\dot
    \psi)^2\bigg)=-\lambda\, e^{B+V}\, e^{a\phi} \,
  \delta(r)\,,\\
}
For  the off-diagonal $\delta G^{ry}$ we get
\eq{
  \label{eomyr}
  &-(D-2)  A' \dot U  +(D-2)  B' \dot U +(D-2)  A' \dot V -{\textstyle \frac 12} \chi' \dot\psi=0  \,.
}
Finally, the dilaton equation of motion becomes
\eq{
  \label{eomdila}
     \Big(\chi''+(D-2) A'\chi'\Big)+\Big(\ddot\psi +(D-2)  \dot U \dot
       \psi\Big)=2a\lambda\, e^{B+V}\, e^{a \phi} \, \delta(r)\,.
}
One first solves these equations in the bulk and then implements the $\delta$-source via a jump of the first derivatives $A', B', \chi'$ at $r=0$. 
We proceed completely analogously to \cite{Blumenhagen:2000dc}, so
that we can keep the presentation short.

\subsubsection*{Solution: type A} 

The first type of solution to these equations of motion are the
direct generalization of the Solution II from \cite{Blumenhagen:2000dc,Blumenhagen:2022mqw} to arbitrary
dimension $D$ and arbitrary coefficient $a$.
The equations in the bulk still admit three free parameters, $\alpha, K, R$, which are further restricted by  implementing the $(D-2)$-brane boundary conditions at $r=0$.
Eventually, we find that the $r$-dependent solutions satisfying
the proper jump conditions at $r=0$ are 
\eq{
 \label{sol2r} 
        A(r)&= \frac{1}{(D-2)} \log\Big\vert \sin\left[ {\textstyle {(D-2) K}} (|r|
          -{\textstyle {\frac R2}})\right] \Big\vert\,,\\[0.2cm]
        \chi(r)&={\frac{\alpha^\pm}{(D-2)}} \log\Big\vert \sin\left[ {\textstyle {(D-2) K}} (|r|
          -{\textstyle {\frac R2}})\right] \Big\vert \\
        &\phantom{aaaaaaaaaaaaaa}\mp 2 \log\Big\vert
        \tan\left[ {\textstyle {{(D-2)\over 2} K}} (|r|
          -{\textstyle {\frac R2}})\right] \Big\vert + \phi_0\,,\\[0.2cm]
         B(r)&={\frac{\mu(\alpha^\pm)}{(D-2)}} \log\Big\vert \sin\left[ {\textstyle {(D-2) K}} (|r|
           -{\textstyle {\frac R2}})\right] \Big\vert \\
         &\phantom{aaaaaaaaaaaaaa}\mp
        {\frac{\alpha^\pm}{(D-2)}} \log\Big\vert \tan\left[ {\textstyle
            {(D-2)\over 2} K} (|r|
          -{\textstyle {\frac R2}})\right] \Big\vert \,,
      }
with  $r\in[-{\frac R2},{\frac R2}]$  and  $\alpha^\pm$ defined below
in \eqref{alpharoots}. Moreover, the coefficient in $B(r)$ reads  
\eq{
  \label{defmu}
  \mu(\alpha)={\alpha^2\over 4(D-2)}+{(D-1)\over 2}\,.
}
The solutions for the $y$-dependent functions are a bit simpler and read
\eq{
            U(y)&=\frac{1}{(D-2)} \log\Big( \cosh\left[ (D-2)K \, y \right] \Big) \,, \\[0.2cm]
              \psi(y)&=\alpha^\pm\, U(y)\,,\quad
        V(y)= \nu(\alpha^\pm)\, U(y)\,,
      }
with
\eq{
  \label{defnu}
  \nu(\alpha)={\alpha^2\over 4(D-2)}-{(D-3)\over 2}\,.
}
Notice that all three functions are proportional and that we have
chosen one integration constant such that the solution is symmetric around $y=0$.

Note that this bulk solution has the free parameter $\alpha$. Imposing
now the necessary boundary condition
\eq{
  \label{hannover96}
  \nu(\alpha)+a\alpha=0\,,
}  
relates $\alpha$  to the parameter $a$ in the brane action via
\eq{ 
     \alpha^2+4a (D-2)\alpha -2(D-2)(D-3)=0\,
}
leading to the two roots
\eq{
 \label{alpharoots} 
    \alpha^\pm=-2a(D-2)\pm 2(D-2)\sqrt{ a^2+{\textstyle {1\over 2}{(D-3)\over (D-2)}}}\,.
  }
 Moreover, consistency of the boundary conditions
 implies\footnote{The dilaton integration constant  $\phi_0$ can be inferred
   from the final boundary condition $e^{a\phi_0}=-{(D-2)\over\lambda} \Delta A'\vert_{r=0}\,e^{-B-a(\chi-\phi_0)}\vert_{r=0}$.}
\eq{
  \label{cosconstraint}
  \cos\left({\textstyle {(D-2)\over 2}}KR\right)={1\over
    \sqrt{a^2+{1\over 2}{(D-3)\over (D-2)}}}\,,
}
which only  admits a solution  $K\sim R^{-1}$ for
\eq{
  \label{regimetypeA}
  |a|\ge \sqrt{(D-1)\over 2(D-2)}\,.
}
Hence, as in the DM example also here a critical value $a_{\rm cr}$
appears. At this value, we get e.g.  $K=0$ where the
$y$-direction trivializes.

In this regime, one can readily compute the size of the $y$-direction,
which comes out finite for $\nu(\alpha)<0$, which means that we
choose the sign in \eqref{alpharoots} such that
$a\alpha>0$, i.e ${\rm sign}(\alpha)={\rm sign}(a)$.
This condition restricts the value of $\alpha$ such that
$|\alpha^\mp|\le \sqrt{2(D-2)(D-3)}$.

Note that  the positive tension enforces a singularity at $r=\pm R/2$,
so that one might expect  to find  $(D-2)$-dimensional
end-of-the-world branes  localized in the $(r,y)$ directions
at $(r,y)=(\pm R/2,\pm \infty)$. However, since the scale factor $e^{B(r)}$
vanishes at $r=\pm R/2$ and the scale factor $e^{V(y)}$ at
$y=\pm\infty$ these points all have zero
distance to each other and in the warped geometry are actually the same point.
For $a$ positive/negative, the scaling of the $y$-distance
to the ETW boundary with the dilaton reads
\eq{
  \label{deltaBFgen}
  \Delta\sim \exp(-\sqrt{2} a {\cal D})\, \qquad{\rm with}\quad {\cal D}\to
  \pm \infty
}  
so that $\delta=2\sqrt{2} |a|$. Using the same notation as in the
DM case, to close-off codimension two singularities we need
ETW-brane of the types
\eq{
  a>a_{\rm cr}:\ {\rm ETW}^{(+)}_{\delta=2\sqrt{2} a}\,,\qquad\quad
   a<-a_{\rm cr}:\ {\rm ETW}^{(-)}_{\delta=-2\sqrt{2} a}\,.
}

\subsubsection*{Solution: type B} 

The question is whether there exist similar solutions in the
regime  $|a|<a_{\rm cr}$, as well. For instance from the relation
\eqref{cosconstraint},
one can get the idea that for these the role of trigonometric 
and hyperbolic functions get exchanged. Indeed, for a positive tension
$(D-1)$-dimensional  brane one finds the bulk solution for the
now hyperbolic $r$-dependent functions
\eq{
 \label{sol2rB} 
        A(r)&= \frac{1}{(D-2)} \log\Big\vert \sinh\left[ {\textstyle {(D-2) K}} (|r|
          -{\textstyle {\frac R2}})\right] \Big\vert\,,\\[0.2cm]
        \chi(r)&={\frac{\alpha^\pm}{(D-2)}} \log\Big\vert \sinh\left[ {\textstyle {(D-2) K}} (|r|
          -{\textstyle {\frac R2}})\right] \Big\vert \\
        &\phantom{aaaaaaaaaaaaaa}\mp 2 \log\Big\vert
        \tanh\left[ {\textstyle {{(D-2)\over 2} K}} (|r|
          -{\textstyle {\frac R2}})\right] \Big\vert + \phi_0\,,\\[0.2cm]
         B(r)&={\frac{\mu(\alpha^\pm)}{(D-2)}} \log\Big\vert \sinh\left[ {\textstyle {(D-2) K}} (|r|
           -{\textstyle {\frac R2}})\right] \Big\vert \\
         &\phantom{aaaaaaaaaaaaaa}\mp
        {\frac{\alpha^\pm}{(D-2)}} \log\Big\vert \tanh\left[ {\textstyle
            {(D-2)\over 2} K} (|r|
          -{\textstyle {\frac R2}})\right] \Big\vert \,,
      }
which feature  a singularity at $|r|=R/2$. Hence we restrict the
regime for the coordinate $r$ to the interval $-R/2\le r\le R/2$.
The now trigonometric functions read
\eq{
            U(y)&=\frac{1}{(D-2)} \log\Big( \cos\left[ (D-2)K \, y \right] \Big) \,, \\[0.2cm]
              \psi(y)&=\alpha^{\pm}\, U(y)\,,\quad
        V(y)= \nu(\alpha^{\pm})\, U(y)\,
}      
with the same
$\mu(\alpha)$ and $\nu(\alpha)$ as in \eqref{defmu} and \eqref{defnu},
respectively. These also have a singularity at $|y|=\pi/(2(D-2)K)$ so
that we restrict the $y$-coordinate to the corresponding interval. 

As for the  former type A solution,  imposing the boundary condition
implies $\nu(\alpha)+\alpha a=0$, which again leads to the two roots in
\eqref{alpharoots}. Now, consistency of the boundary condition implies 
\eq{
  \cosh\left(\frac{(D-2)}{2}KR\right)=\frac{1}{\sqrt{a^2+\frac{1}{2}\frac{(D-3)}{(D-2)}}}\,,}
which admits a solution in the regime
\eq{
  |a|\leq\sqrt{\frac{(D-1)}{2(D-2)}}
}
that is complementary to the regime \eqref{regimetypeA}.
We find that always both
the proper size of the $r$ and the $y$-interval turns out to be
finite.

\begin{figure}[ht]
  \centering
   \includegraphics[width=9.0cm]{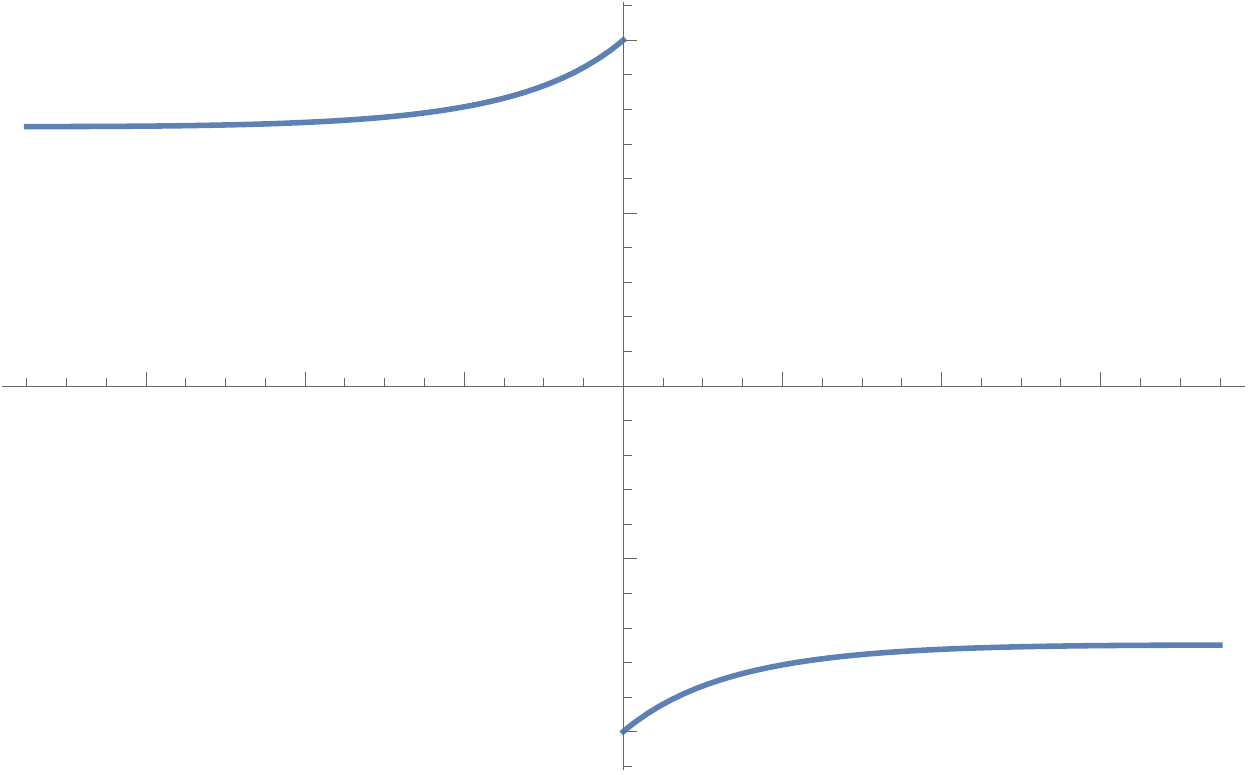}
  \begin{picture}(0,0)
    \put(-19,86){$a_{\rm cr}$}
     \put(-270,67){$-a_{\rm cr}$}
    \put(-134,162){$\varepsilon$}
    \put(-144,6){\tiny $-1$}
    \put(-129,149){\tiny$1$}
    \put(-137,130){$-a_{\rm cr}$}
    \put(-160,23){$-a_{\rm cr}-$}
    \put(-11,78){\tiny $\vert$}
    \put(-256,78){\tiny $\vert$}
    \end{picture}
    \caption{The scaling parameter  $\varepsilon$ as a function of $a$
      in the regime $|a|\le a_{\rm cr}$.}
  \label{fig:deltaB}
\end{figure}
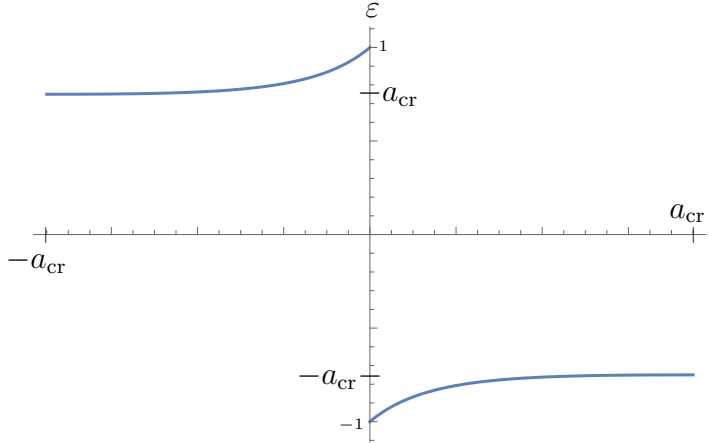

We seem to have singularities at the four points
$(r,y)=(\pm {R\over 2},\pm{\pi\over 2K(D-2)})$  whose distances
and the induced singularity structure depends on the value of
$\alpha$. Hence, a complete analysis now requires a more
intricate analysis, which is beyond the scope of this paper.
Thus, we now simply choose the branch  $a\alpha>0$ and compute
the scaling behavior 
when approaching the singularity at $r=R/2$ along
the $r$-direction.  We find 
\eq{
  \label{scaletypeB}
  \Delta\sim \exp(\varepsilon \,\chi)\,\quad {\rm with}\qquad
  \varepsilon={\mu(\alpha^\pm)\mp \alpha^\pm +(D-2)
    \over \alpha^\pm \mp 2(D-2)}\,.
}
Clearly this looks more complicated than \eqref{deltaBFgen} for the type A model,
but as shown in figure \ref{fig:deltaB},  in the regime
of interest $|a|\le a_{\rm cr}$, $\varepsilon$ is monotonically increasing
with $a$ and it is  bounded by 
\eq{
       0\le a\le a_{\rm cr}:&\qquad   -1\le \varepsilon\le -a_{\rm cr}\\
       -a_{\rm cr}\le a\le 0:&\qquad\ \;   a_{\rm cr}\le \varepsilon\le
       1\,.
     }
It is very remarkable that precisely the critical  value $a_{\rm cr}$ appears
here.

%\vspace{0.3cm}
%%%%%%%%%%%%
%%%%%%%%%%%%

Taking also the scaling of the dilaton into account, we expect this
singular type B solution to be closed off by $(D-2)$-dimensional
ETW-branes of the types
\eq{
 0< a< a_{\rm cr}:\ {\rm ETW}^{(+)}_{\delta=-2\sqrt{2}\, \varepsilon}\,,\qquad\quad
  -a_{\rm cr}\le a < 0:\ {\rm ETW}^{(-)}_{\delta=2\sqrt{2}\, \varepsilon}\,.
 }
 Note that both for the type A and the type B  solution we get
 $\delta\ge 2\sqrt{2}\, a_{\rm cr}$.

\subsection{BF ETW-brane  solution}
\label{sec:BFETW}

Following the Dynamical Cobordism Conjecture, we expect the existence
of  explicit solutions for the $(D-2)$-dimensional end-of-the world
branes. These are again expected to preserve the longitudinal 
$(D-2)$-dimensional  Poincar\'e invariance, but break the transversal
rotational symmetry.
Thus,  we consider the action of a codimension two defect brane in
$D$-dimensions
\eq{
  \label{actionetw}
        S={1\over 2}\int d^{D}x  \sqrt{-G} \left( R-{1\over 2} (\partial
          {\phi})^2 \right) - \hat{\lambda} \int d^{D}x \sqrt{-g}\, e^{\hat{a}{\phi}}\,
        \delta^2(y)\,.
      }
The ten-dimensional case with $\hat a=5/4$  was analysed recently in
\cite{Blumenhagen:2022mqw}. It turns out that the general case can be
studied in a very similar
fashion, so that we keep the following presentation short and to the point.

For the $D$-dimensional metric we make the non-isotropic ansatz
\eq{
        ds^2=e^{2{\mathcal{A}}(\rho,\varphi)}ds_{D-2}^2+e^{2{\mathcal{B}}(\rho,\varphi)}(d\rho^2+\rho^2d\varphi^2)
      }
with a separated dependence of the warp factor
${\mathcal{A}},{\mathcal{B}}$ and the dilaton $\phi$ on the radial and
angular coordinates $\rho$ and $\varphi$, i.e.
\eq{
&\mathcal{A}(\rho,\varphi)=A(\rho)+U(\varphi), \,\,\,\mathcal{B}(\rho,\varphi)=B(\rho)+V(\varphi), \\
&\phi(\rho,\varphi)=\chi(\rho)+\psi(\varphi).
}
The resulting equations of motion read
\begin{eqnarray}
\label{eometw}
    &&\hspace{-0.5cm}\mathbf{\delta G^{\mu\nu}}:\quad
   \bigg((D-3) {A}^{\prime\prime}+(D-3)\frac{{A}^{\prime}}{\rho}+{\textstyle\frac{(D-2)(D-3)}{2}}({A}^{\prime})^2+{B}^{\prime\prime}+\frac{{B}^{\prime}}{\rho}+\frac{1}{4}({\chi}^{\prime})^2\bigg)\nonumber\\
    &&\hspace{-0.5cm}\phantom{aaii} +\frac{1}{\rho^2}\bigg((D-3)\ddot{{U}}+{\textstyle\frac{(D-2)(D-3)}{2}}(\dot{{U}})^2+\ddot{{V}}+\frac{1}{4}(\dot{{\psi}})^2\bigg)
     =-\hat{\lambda}\,  e^{\hat{a}
       {\phi}}\,\frac{\delta(\rho)}{2\pi\rho}\nonumber\\[0.2cm]    
  &&\hspace{-0.5cm}  \mathbf{\delta G^{\rho\rho}}:\quad
    \bigg((D-2)\frac{{A}^{\prime}}{\rho}+{\textstyle \frac{(D-2)(D-3)}{2}}({A}^{\prime})^2+(D-2){A}^{\prime}{B}^{\prime}-\frac{1}{4}({\chi}^{\prime})^2\bigg)\nonumber\\
     &&\hspace{-0.5cm}\phantom{aaa}  +\frac{1}{\rho^2}\bigg((D-2)\ddot{{U}}+{\textstyle \frac{(D-2)(D-1)}{2}}(\dot{{U}})^2-(D-2)\dot{U}\dot{{V}}+\frac{1}{4}(\dot{{\psi}})^2\bigg)
     =0\\[0.2cm] 
  &&\hspace{-0.5cm}  \mathbf{\delta G^{\varphi\varphi}}:\quad
    \bigg((D-2){A}^{\prime\prime}+{\textstyle \frac{(D-2)(D-1)}{2}}({A}^{\prime})^2-(D-2){A}^{\prime}{B}^{\prime}+\frac{1}{4}({\chi}^{\prime})^2\bigg)\nonumber\\
     &&\hspace{-0.5cm}\phantom{aaai}  +\frac{1}{\rho^2}\bigg({\textstyle \frac{(D-2)(D-3)}{2}}(\dot{{U}})^2+(D-2)\dot{{U}}\dot{{V}}-\frac{1}{4}(\dot{{\psi}})^2\bigg)
     =0\nonumber\\[0.2cm] 
   &&\hspace{-0.5cm} \mathbf{\delta G^{\rho\varphi}}:\quad
     (D-2)\frac{\dot{{U}}}{\rho}-(D-2){A}^{\prime}\dot{{U}}+(D-2){B}^{\prime}\dot{{U}}+(D-2){A}^{\prime}\dot{{V}}-\frac{1}{2}{\chi}^{\prime}\dot{{\psi}}=0\nonumber\\[0.2cm] 
    && \hspace{-0.5cm}\mathbf{\delta\phi}:\quad 
     \bigg({\chi}^{\prime\prime}+\frac{{\chi}^{\prime}}{\rho}+(D-2){A}^{\prime}{\chi}^{\prime}\bigg)+\frac{1}{\rho^2}\bigg(\ddot{{\psi}}+(D-2)\dot{{U}}\dot{{\psi}}\bigg)=2\hat{a}\hat{\lambda}\,
     e^{\hat{a} {\phi}}\,\frac{\delta(\rho)}{2\pi\rho}\,.\nonumber
\end{eqnarray}

We are seeking for solutions to these equations that close to 
specific points feature the same scaling as the type A and type B
neutral brane backreactions from the previous section.
We proceed analogously to  \cite{Blumenhagen:2022mqw} and would like
to refer the reader to that paper for more details on the behavior
and interpretation of the solution. The only difference
is that here we are consistently working
in Einstein-frame, whereas certain quantities were computed in
string frame in \cite{Blumenhagen:2022mqw}.

\subsubsection*{ETW-brane: type A}

Generalizing the result from \cite{Blumenhagen:2022mqw}, we find the
$D$-dimensional bulk solution for the radial functions
\eq{
    &{A}(\rho)=\frac{1}{(D-2)}\log\left(\cosh\left[(D-2)\hat{K}\log\left(\frac{\rho}{\rho_0}\right)\right]\right)\,,\\[0.1cm]
    &{\chi}(\rho)=\hat{\alpha}{A}(\rho)\,,\\[0.1cm]
    &{B}(\rho)=-\log\left(\frac{\rho}{\rho_0}\right)+\nu(\hat\alpha)\,{A}(\rho)
}
and the angular functions
\eq{
\label{trigonA}  
    &{U}(\varphi)=\frac{1}{(D-2)}\log\left|\cos\left((D-2)\hat{K}\varphi\right)\right|\,,\\[0.1cm]
    &{\psi}(\varphi)=\frac{\hat{\alpha}}{(D-2)}\log\left|\cos\left((D-2)\hat{K}\varphi\right)\right|\\
    &\phantom{aaaaaaaaaaaaaa}\pm2\log\left|\tan\left(\frac{(D-2)}{2}\hat{K}\varphi+\frac{\pi}{4}\right)\right|\,,\\[0.1cm]
    &{V}(\varphi)={\mu(\hat\alpha)\over
        (D-2)}\log\left|\cos\left((D-2)\hat{K}\varphi\right)\right|\\
&\phantom{aaaaaaaaaaaaaa}\pm {\hat\alpha\over (D-2)}\log\left|\tan\left(\frac{(D-2)}{2}\hat{K}\varphi+\frac{\pi}{4}\right)\right|\,.
} 
Here $\hat\alpha$ and $\hat K$ are still free parameters.
Imposing now the boundary condition, following the same steps as in
\cite{Blumenhagen:2022mqw}, we get $\hat a=0$ and $\hat{\lambda}=2\pi$
so that the  ETW $(D-3)$-brane action in
Einstein-frame action reads 
\eq{
  S_{\rm ETW}=-2\pi \int{d^Dx\,\sqrt{-g}\,\,\frac{\delta(\rho)}{2\pi\rho}}\,.
}

Next, we analyze the behavior of the solution close to the core at
$\rho=0$ and compute the distance
\eq{
  \Delta\sim\int_0^{\rho}d\rho'\, e^{{B}(\rho')}
  \sim\rho^{-\hat{K}\left(\frac{\hat{\alpha}^2}{4(D-2)}-\frac{(D-3)}{2}\right)}
  \sim \rho^{-\hat K \,\nu(\hat \alpha)}
}
where the function $\nu(\hat\alpha)$ has been defined in
\eqref{defnu}. Note that we have implicitly used that $\nu(\hat\alpha)$
is actually negative for the distance to be finite, a point 
that will become clear below.
Close to the core, the dilaton behaves as $\chi(\rho)\sim -\hat K
\hat\alpha\log\rho$, so that we find the scaling
\eq{
     \Delta\sim \exp\left(\sqrt{2}\, {\nu(\hat\alpha)\over \hat\alpha}\, {\cal D}\right)\,.
   }
Comparing this to the scaling of the distance in the type A solution
\eqref{deltaBFgen}, we can read off
  $a=-\nu(\hat\alpha)/\hat\alpha$. Taking into account the boundary condition
  \eqref{hannover96}, it is clear that this relation is solved
  for
  \eq{
    \hat\alpha = \alpha\,.
  }
 Consistently, the condition $a\alpha>0$ (from section \ref{sec:genBFsol}) and the
 relation \eqref{hannover96} implies
 $\nu(\hat\alpha)<0$ so that the distance to the boundary is indeed finite
 in both set-ups simultaneously.

Recalling that for the type A solution we found  the lower bound
\eqref{cosconstraint} on the parameter $a=-\nu(\hat\alpha)/\hat\alpha$, one might wonder how such a similar
bound could arise directly in the ETW-brane solution.
To develop an idea, we first  observe that in the type A solution the initial codimension
one brane reaches the singularity. Hence, for the ETW-brane to really
close-off the singularity of the type A solution, this brane must also
be included in its solution. This can be done by adding
to the action \eqref{actionetw} a source term
\eq{
                  S=-\lambda \int d^Dx\, \sqrt{ -g }\,e^{a\phi}\,
                  {\delta(\varphi)\over \rho}\,
                }
describing the codimension
one brane stretched along the $\rho$ direction at fixed value of
$\varphi$. Here the tension $\lambda$ and the parameter $a$
are really the ones from the action \eqref{actionBFgen}.

One realizes that this additional boundary condition
on the right hand side of the equations of motion \eqref{eometw}
can straightforwardly be satisfied by the bulk solution 
if we  replace in all the trigonometric functions in \eqref{trigonA}
\eq{
       (D-2) \hat K \varphi \to  (D-2) \hat K \big(|\varphi|+\hat L\big)\,.
}          
From here on the computation is completely analogous to imposing
the boundary conditions for the type A solution. Hence we get for
instance the condition $\nu(\hat\alpha)+a \hat\alpha$ leading
to the constraint
\eq{
    \label{cosconstraintetw}
  \cos\Big({\textstyle {(D-2)}}\hat K \hat L\Big)={1\over
    \sqrt{a^2+{1\over 2}{(D-3)\over (D-2)}}}\,
}
and consequently to the lower bound $|a|=|-\nu(\hat\alpha)/\hat\alpha|\ge a_{\rm cr}$.

\begin{figure}[ht]
  \centering
   \includegraphics[width=8.0cm]{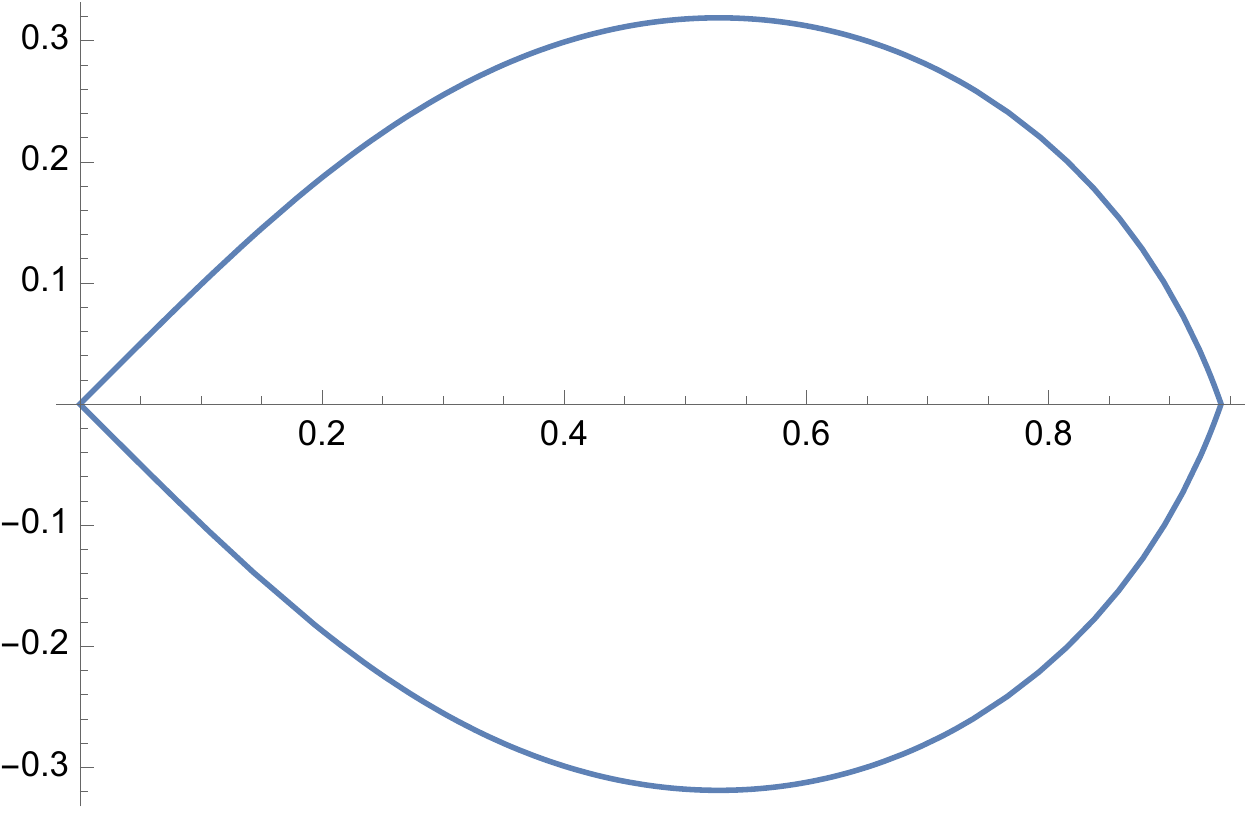}
  \begin{picture}(0,0)
    \end{picture}
    \caption{Polar-plot of the warp factor $e^{V(\varphi)}$ for $D=10$,
      $a=5/4$, $\hat K=1/8$ and $L=\pi/4$ in the regime $-\pi/4\le
      \varphi\le \pi/4$.}
  \label{fig:polarpl}
\end{figure}

In figure \ref{fig:polarpl} we show the behavior of the resulting warp factor
$e^{V(\varphi)}$ in the relevant regime $|\varphi|\le {\pi\over
  2(D-2)\hat K}-\hat L$, where we think of the two boundaries to be
identified. One clearly sees the kink induced by the
boundary conditions for the codimension one brane at $\varphi=0$. Moreover,
at the boundary the length scale in the $\rho$ direction goes to zero
and space disappears.

%\vspace{0.3cm}
%%%%%%%%%%%%
%%%%%%%%%%%%
  
 \subsubsection*{ETW-brane: type B}

 Very similarly one can find a solution to the  bulk equations of motion
 \eqref{eometw}  that carries  the features to  close-off the type B solution, which exists
 for $|a|\le a_{\rm cr}$. The radial functions read
 \eq{
   \label{radialsoletwB}
    &{A}(\rho)=\frac{1}{(D-2)}\log\left\vert\sinh\left[(D-2)\hat{K}\log\left(\frac{\rho}{\rho_0}\right)\right]\right\vert\,,\\
    &{\chi}(\rho)=\hat{\alpha}{A}(\rho)\pm 2
\log\left\vert\coth\left[{(D-2)\over 2}\hat{K}\log\left(\frac{\rho}{\rho_0}\right)\right]\right\vert
    \,,\\
    &{B}(\rho)=-\log\left(\frac{\rho}{\rho_0}\right)+\mu(\hat\alpha)\,{A}(\rho)\\
     &\phantom{aaaaaaaa}\pm {\hat\alpha\over (D-2)}
\log\left\vert\coth\left[{(D-2)\over 2}\hat{K}\log\left(\frac{\rho}{\rho_0}\right)\right]\right\vert
}
and the angular functions
\eq{   
    &{U}(\varphi)=\frac{1}{(D-2)}\log\left|\cos\left((D-2)\hat{K}\varphi\right)\right|\,,\\[0.1cm]
    &{\psi}(\varphi)=\hat\alpha\, {U}(\varphi)\,,\qquad
     {V}(\varphi)=\nu(\hat\alpha) \,{U}(\varphi)\,.
   }
As for the type B solution from the previous section, the
implementation of boundary conditions is more intricate than for the
type A ETW-brane solution so that again we restrict ourselves to discuss only
the for this discussion most important aspect and leave a more
thorough analysis to future work.
We observe that close to $\rho=\rho_0$ this solution shows the
scaling behavior
\eq{
                \Delta=\int_{\rho_0}^\rho  d\rho' e^{B(\rho')}\sim
                \int_0^x dx'\, (x')^{\mu\mp\hat\alpha\over D-2}\sim
                x^{\mu\mp\hat\alpha+(D-2)\over D-2}
  }
where  we have substituted $\rho=\rho_0+\rho_0 x$ and expanded for
small $x$.   Moreover, the dilaton behaves as
\eq{
  \chi(x)\sim {\hat\alpha\mp 2(D-2)\over (D-2)} \log x
}  
so that we get the scaling behavior
\eq{
  \Delta\sim \exp(\varepsilon \,\chi)\,\quad {\rm with}\qquad
  \varepsilon={\mu(\hat\alpha)\mp \hat\alpha +(D-2)
    \over \hat\alpha \mp 2(D-2)}\,,
}
which  for $\hat\alpha=\alpha$ indeed  agrees with \eqref{scaletypeB} for the solution of type
B.

 \subsection{Dynamical Cobordism Conjecture: BF case}

In view of the  Dynamical  Cobordism Conjecture
we can now summarize:
\begin{quotation}
\noindent{\it The neutral $(D-2)$-dimensional ETW-brane of type A(B) can close-off the
  singular   BF solution of type A(B)  with
  $|a|> a_{\rm cr}$ {\rm (}$|a|< a_{\rm cr}${\rm )}.  The scaling parameter $\delta$
  is bounded from below by $\delta \ge \delta_{\rm cr}$ with the same
  critical value as in the DM case.}
\end{quotation}
 
\noindent
One can see that for the critical case $|a|=a_{\rm cr}$ the longitudinal $y$-direction
trivializes (for $K=0$) and that the solution goes over to a  solution
of the form discussed for  the neutral DM ETW-brane in section
\ref{sec:nonbps}. Indeed, both initial actions are the same and
$a_{\rm cr}=|a_0|$. 

Recall that in the generalized  BF-model, we were constructing
the required codimension  two ETW-branes entirely in dilaton gravity,
i.e. without any coupling to additional higher form fields.
Hence, there are no charged ETW-branes required, which is consistent
with the fact that the initial BF-action \eqref{actionBFgen} is not generated by some
background flux.

Finally, we note that in string-frame the  ETW 7-brane action in ten dimensions
takes the form
\eq{
  S_{\rm ETW}=-2\pi \int{d^{10} x\,\sqrt{-g}\,\,e^{-2\phi}\,\frac{\delta(\rho)}{2\pi\rho}}\,,
}
which is not a familiar  brane action. The dilaton coupling is rather reminiscent of the
action of the solitonic NS-NS 5-brane.

\section{Conclusions}

We have analyzed simple $D$-dimensional models featuring
the phenomenon of dynamical cobordism from the perspective
of the required end-of-the-world defects to close off the
singular solutions. We were invoking a  Dynamical Cobordism
Conjecture for on-shell configurations consistent with quantum gravity.
It says that, if we have an on-shell solution with a singularity at finite distance, then
the required codimension one ETW-branes
must also be local on-shell solutions of the respective equations of
motion.

For a generalized Dudas-Mourad model, we constructed both  neutral and
charged ETW-brane solutions, showing precisely the expected scaling behavior close
to their respective cores. Moreover, in that regime the dilaton
and the warp factor of the DM like solutions agree with the ones of the backreacted
ETW-brane solutions when expressed in
terms of the proper field distance. Hence, the ETW-brane boundary conditions can smoothly
be implemented in the DM like solutions which is what one means by
the neutral and charged ETW-branes to 
closed-off the DM like solutions.

Closing-off the solution by only   neutral ETW-branes turned out to be
only possible for the parameter $|c|$ being below a critical value.
For values of $|c|$ above this bound, also  charged ETW-branes
are needed to  close off the DM solution. 
Moreover, it was shown that both the  rolling and the
ETW-brane solutions could be made consistent with
dimensional reduction.
Since our solutions exhibit an intriguing lower bound on the scaling
parameter $\delta$,  together with the bound from the
sharpened Swampland Distance Conjecture we were deriving an upper bound
on the scaling parameter in a generalized de Sitter Distance
Conjecture.

We also extended our  systematic analysis  to the solution of
a generalized Blumenhagen-Font \cite{Blumenhagen:2000dc} model.
In this case, a critical value $a_{\rm cr}$ appeared, as well,
separating two  different solutions to the equations of motion, dubbed
type A and type B.
We have argued  that both these solutions can be closed-off
by corresponding  non-isotropic  ETW-branes.

Our analysis gives credence to the idea that such finite size
running solutions induced by dynamical tadpoles
feature finite distance singularities that 
can be closed-off by corresponding ETW-branes.
The examples discussed in this paper support the picture
that also the ETW-branes can be explicitly described
by on-shell solutions to the equations of motion.
Of course, in general such
non-supersymmetric solutions could  contain tachyonic
instabilities, though this is an issue of a different kind, which would
certainly be  interesting to investigate.

In this paper, we discussed ETW-branes of codimension one and two
and it would be interesting whether one can also find explicit
solutions for dynamical tadpoles and the corresponding ETW-branes 
in higher codimension. 
Moreover, one expects
that by exchanging the dynamical space-like direction
with the time-like one, one could also
generalize some of the solutions presented here to
cosmological ones \cite{Angius:2022mgh, Dudas:2010gi, Mourad:2021roa}.

%\vspace{0.2cm}

\noindent
\paragraph{Acknowledgments:}
We would like to thank Ivano Basile, Niccolo Cribiori and Andriana Makridou for useful discussions.

\vspace{0.4cm}
\clearpage

\bibliography{references}  

\providecommand{\href}[2]{#2}\begingroup\raggedright\begin{thebibliography}{10}

\bibitem{Banks:2010zn}
T.~Banks and N.~Seiberg, ``{Symmetries and Strings in Field Theory and
  Gravity},'' {\em Phys. Rev. D} {\bf 83} (2011) 084019,
  \href{http://www.arXiv.org/abs/1011.5120}{{\tt 1011.5120}}.

\bibitem{Banks:1988yz}
T.~Banks and L.~J. Dixon, ``{Constraints on String Vacua with Space-Time
  Supersymmetry},'' {\em Nucl. Phys. B} {\bf 307} (1988) 93--108.

\bibitem{Palti:2019pca}
E.~Palti, ``{The Swampland: Introduction and Review},'' {\em Fortsch. Phys.}
  {\bf 67} (2019), no.~6, 1900037,
  \href{http://www.arXiv.org/abs/1903.06239}{{\tt 1903.06239}}.

\bibitem{vanBeest:2021lhn}
M.~van Beest, J.~Calder\'on-Infante, D.~Mirfendereski, and I.~Valenzuela,
  ``{Lectures on the Swampland Program in String Compactifications},''
  \href{http://www.arXiv.org/abs/2102.01111}{{\tt 2102.01111}}.

\bibitem{Grana:2021zvf}
M.~Gra\~na and A.~Herr\'aez, ``{The Swampland Conjectures: A Bridge from
  Quantum Gravity to Particle Physics},'' {\em Universe} {\bf 7} (2021), no.~8,
  273, \href{http://www.arXiv.org/abs/2107.00087}{{\tt 2107.00087}}.

\bibitem{McNamara:2019rup}
J.~McNamara and C.~Vafa, ``{Cobordism Classes and the Swampland},''
  \href{http://www.arXiv.org/abs/1909.10355}{{\tt 1909.10355}}.

\bibitem{GarciaEtxebarria:2020xsr}
I.~Garc\'\i{}a~Etxebarria, M.~Montero, K.~Sousa, and I.~Valenzuela, ``{Nothing
  is certain in string compactifications},'' {\em JHEP} {\bf 12} (2020) 032,
  \href{http://www.arXiv.org/abs/2005.06494}{{\tt 2005.06494}}.

\bibitem{Montero:2020icj}
M.~Montero and C.~Vafa, ``{Cobordism Conjecture, Anomalies, and the String
  Lamppost Principle},'' {\em JHEP} {\bf 01} (2021) 063,
  \href{http://www.arXiv.org/abs/2008.11729}{{\tt 2008.11729}}.

\bibitem{Dierigl:2020lai}
M.~Dierigl and J.~J. Heckman, ``{Swampland cobordism conjecture and non-Abelian
  duality groups},'' {\em Phys. Rev. D} {\bf 103} (2021), no.~6, 066006,
  \href{http://www.arXiv.org/abs/2012.00013}{{\tt 2012.00013}}.

\bibitem{Hamada:2021bbz}
Y.~Hamada and C.~Vafa, ``{8d supergravity, reconstruction of internal geometry
  and the Swampland},'' {\em JHEP} {\bf 06} (2021) 178,
  \href{http://www.arXiv.org/abs/2104.05724}{{\tt 2104.05724}}.

\bibitem{Debray:2021vob}
A.~Debray, M.~Dierigl, J.~J. Heckman, and M.~Montero, ``{The anomaly that was
  not meant IIB},'' \href{http://www.arXiv.org/abs/2107.14227}{{\tt
  2107.14227}}.

\bibitem{McNamara:2021cuo}
J.~McNamara, ``{Gravitational Solitons and Completeness},''
  \href{http://www.arXiv.org/abs/2108.02228}{{\tt 2108.02228}}.

\bibitem{Blumenhagen:2021nmi}
R.~Blumenhagen and N.~Cribiori, ``{Open-closed correspondence of K-theory and
  cobordism},'' {\em JHEP} {\bf 08} (2022) 037,
  \href{http://www.arXiv.org/abs/2112.07678}{{\tt 2112.07678}}.

\bibitem{Andriot:2022mri}
D.~Andriot, N.~Carqueville, and N.~Cribiori, ``{Looking for structure in the
  cobordism conjecture},'' {\em SciPost Phys.} {\bf 13} (2022), no.~3, 071,
  \href{http://www.arXiv.org/abs/2204.00021}{{\tt 2204.00021}}.

\bibitem{Blumenhagen:2022bvh}
R.~Blumenhagen, N.~Cribiori, C.~Kneissl, and A.~Makridou, ``{Dimensional
  Reduction of Cobordism and K-theory},''
  \href{http://www.arXiv.org/abs/2208.01656}{{\tt 2208.01656}}.

\bibitem{Velazquez:2022eco}
D.~M. Vel\'azquez, D.~De~Biasio, and D.~L{\"u}st, ``{Cobordism, Singularities
  and the Ricci Flow Conjecture},''
  \href{http://www.arXiv.org/abs/2209.10297}{{\tt 2209.10297}}.

\bibitem{McNamara:2022lrw}
J.~McNamara and M.~Reece, ``{Reflections on Parity Breaking},''
  \href{http://www.arXiv.org/abs/2212.00039}{{\tt 2212.00039}}.

\bibitem{Dierigl:2022reg}
M.~Dierigl, J.~J. Heckman, M.~Montero, and E.~Torres, ``{IIB Explored:
  Reflection 7-Branes},'' \href{http://www.arXiv.org/abs/2212.05077}{{\tt
  2212.05077}}.

\bibitem{Debray:2023yrs}
A.~Debray, M.~Dierigl, J.~J. Heckman, and M.~Montero, ``{The Chronicles of
  IIBordia: Dualities, Bordisms, and the Swampland},''
  \href{http://www.arXiv.org/abs/2302.00007}{{\tt 2302.00007}}.

\bibitem{Buratti:2021yia}
G.~Buratti, M.~Delgado, and A.~M. Uranga, ``{Dynamical tadpoles, stringy
  cobordism, and the SM from spontaneous compactification},'' {\em JHEP} {\bf
  06} (2021) 170, \href{http://www.arXiv.org/abs/2104.02091}{{\tt 2104.02091}}.

\bibitem{Buratti:2021fiv}
G.~Buratti, J.~Calder\'on-Infante, M.~Delgado, and A.~M. Uranga, ``{Dynamical
  Cobordism and Swampland Distance Conjectures},'' {\em JHEP} {\bf 10} (2021)
  037, \href{http://www.arXiv.org/abs/2107.09098}{{\tt 2107.09098}}.

\bibitem{Angius:2022aeq}
R.~Angius, J.~Calder\'on-Infante, M.~Delgado, J.~Huertas, and A.~M. Uranga,
  ``{At the end of the world: Local Dynamical Cobordism},'' {\em JHEP} {\bf 06}
  (2022) 142, \href{http://www.arXiv.org/abs/2203.11240}{{\tt 2203.11240}}.

\bibitem{Blumenhagen:2022mqw}
R.~Blumenhagen, N.~Cribiori, C.~Kneissl, and A.~Makridou, ``{Dynamical
  cobordism of a domain wall and its companion defect 7-brane},'' {\em JHEP}
  {\bf 08} (2022) 204, \href{http://www.arXiv.org/abs/2205.09782}{{\tt
  2205.09782}}.

\bibitem{Angius:2022mgh}
R.~Angius, M.~Delgado, and A.~M. Uranga, ``{Dynamical Cobordism and the
  beginning of time: supercritical strings and tachyon condensation},'' {\em
  JHEP} {\bf 08} (2022) 285, \href{http://www.arXiv.org/abs/2207.13108}{{\tt
  2207.13108}}.

\bibitem{Antonelli:2019nar}
R.~Antonelli and I.~Basile, ``{Brane annihilation in non-supersymmetric
  strings},'' {\em JHEP} {\bf 11} (2019) 021,
  \href{http://www.arXiv.org/abs/1908.04352}{{\tt 1908.04352}}.

\bibitem{Blumenhagen:2000dc}
R.~Blumenhagen and A.~Font, ``{Dilaton tadpoles, warped geometries and large
  extra dimensions for nonsupersymmetric strings},'' {\em Nucl. Phys. B} {\bf
  599} (2001) 241--254, \href{http://www.arXiv.org/abs/hep-th/0011269}{{\tt
  hep-th/0011269}}.

\bibitem{Basile:2021vxh}
I.~Basile, ``{Supersymmetry breaking and stability in string vacua: Brane
  dynamics, bubbles and the swampland},'' {\em Riv. Nuovo Cim.} {\bf 44}
  (2021), no.~10, 499--596, \href{http://www.arXiv.org/abs/2107.02814}{{\tt
  2107.02814}}.

\bibitem{Raucci:2022bjw}
S.~Raucci, ``{On new vacua of non-supersymmetric strings},'' {\em Phys. Lett.
  B} {\bf 837} (2023) 137663, \href{http://www.arXiv.org/abs/2209.06537}{{\tt
  2209.06537}}.

\bibitem{Basile:2022ypo}
I.~Basile, S.~Raucci, and S.~Thom\'ee, ``{Revisiting Dudas-Mourad
  Compactifications},'' {\em Universe} {\bf 8} (2022), no.~10, 544,
  \href{http://www.arXiv.org/abs/2209.10553}{{\tt 2209.10553}}.

\bibitem{Raucci:2022jgw}
S.~Raucci, ``{On codimension-one vacua and string theory},'' {\em Nucl. Phys.
  B} {\bf 985} (2022) 116002, \href{http://www.arXiv.org/abs/2206.06399}{{\tt
  2206.06399}}.

\bibitem{Bedroya:2019snp}
A.~Bedroya and C.~Vafa, ``{Trans-Planckian Censorship and the Swampland},''
  {\em JHEP} {\bf 09} (2020) 123,
  \href{http://www.arXiv.org/abs/1909.11063}{{\tt 1909.11063}}.

\bibitem{Etheredge:2022opl}
M.~Etheredge, B.~Heidenreich, S.~Kaya, Y.~Qiu, and T.~Rudelius, ``{Sharpening
  the Distance Conjecture in Diverse Dimensions},''
  \href{http://www.arXiv.org/abs/2206.04063}{{\tt 2206.04063}}.

\bibitem{Lust:2019zwm}
D.~L\"ust, E.~Palti, and C.~Vafa, ``{AdS and the Swampland},'' {\em Phys. Lett.
  B} {\bf 797} (2019) 134867, \href{http://www.arXiv.org/abs/1906.05225}{{\tt
  1906.05225}}.

\bibitem{Sugimoto:1999tx}
S.~Sugimoto, ``{Anomaly cancellations in type I D-9 - anti-D-9 system and the
  USp(32) string theory},'' {\em Prog. Theor. Phys.} {\bf 102} (1999) 685--699,
  \href{http://www.arXiv.org/abs/hep-th/9905159}{{\tt hep-th/9905159}}.

\bibitem{Dudas:2000ff}
E.~Dudas and J.~Mourad, ``{Brane solutions in strings with broken supersymmetry
  and dilaton tadpoles},'' {\em Phys. Lett. B} {\bf 486} (2000) 172--178,
  \href{http://www.arXiv.org/abs/hep-th/0004165}{{\tt hep-th/0004165}}.

\bibitem{Duff:1994an}
M.~J. Duff, R.~R. Khuri, and J.~X. Lu, ``{String solitons},'' {\em Phys. Rept.}
  {\bf 259} (1995) 213--326,
  \href{http://www.arXiv.org/abs/hep-th/9412184}{{\tt hep-th/9412184}}.

\bibitem{Bedroya:2022tbh}
A.~Bedroya, ``{Holographic origin of TCC and the Distance Conjecture},''
  \href{http://www.arXiv.org/abs/2211.09128}{{\tt 2211.09128}}.

\bibitem{Ooguri:2006in}
H.~Ooguri and C.~Vafa, ``{On the Geometry of the String Landscape and the
  Swampland},'' {\em Nucl. Phys. B} {\bf 766} (2007) 21--33,
  \href{http://www.arXiv.org/abs/hep-th/0605264}{{\tt hep-th/0605264}}.

\bibitem{Lee:2019xtm}
S.-J. Lee, W.~Lerche, and T.~Weigand, ``{Emergent strings, duality and weak
  coupling limits for two-form fields},'' {\em JHEP} {\bf 02} (2022) 096,
  \href{http://www.arXiv.org/abs/1904.06344}{{\tt 1904.06344}}.

\bibitem{Lee:2019wij}
S.-J. Lee, W.~Lerche, and T.~Weigand, ``{Emergent strings from infinite
  distance limits},'' {\em JHEP} {\bf 02} (2022) 190,
  \href{http://www.arXiv.org/abs/1910.01135}{{\tt 1910.01135}}.

\bibitem{Higuchi:1986py}
A.~Higuchi, ``{Forbidden Mass Range for Spin-2 Field Theory in De Sitter
  Space-time},'' {\em Nucl. Phys. B} {\bf 282} (1987) 397--436.

\bibitem{Dudas:2010gi}
E.~Dudas, N.~Kitazawa, and A.~Sagnotti, ``{On Climbing Scalars in String
  Theory},'' {\em Phys. Lett. B} {\bf 694} (2011) 80--88,
  \href{http://www.arXiv.org/abs/1009.0874}{{\tt 1009.0874}}.

\bibitem{Mourad:2021roa}
J.~Mourad and A.~Sagnotti, ``{On warped string vacuum profiles and cosmologies.
  Part II. Non-supersymmetric strings},'' {\em JHEP} {\bf 12} (2021) 138,
  \href{http://www.arXiv.org/abs/2109.12328}{{\tt 2109.12328}}.

\end{thebibliography}\endgroup
\bibliographystyle{utphys}

%%%%%%%%%%%%%%%%%%%%%%%%%%%%%%%%%%%%%%%%%%%%%%%
%%%%%%%%%%%%%%%%%%%%%%%%%%%%%%%%%%%%%%%%%%%%%%%
%%%%%%%%%%%%%%%%%%%%%%%%%%%%%%%%%%%%%%%%%%%%%%%
%%%%%%%%%%%%%%%%%%%%%%%%%%%%%%%%%%%%%%%%%%%%%%%

\end{document}